%% file: ANA-IDTR-2021-03-PAPER.tex
\documentclass[coverpage=false, cernpreprint, atlasdraft=false, texlive=2023, UKenglish, texmf, orcidlogo]{atlasdoc}
 
\usepackage{atlaspackage}
\usepackage{atlasbiblatex}
 
\usepackage{atlascontribute}
 
\usepackage{atlasphysics}
\usepackage{xcolor}
\usepackage{float}
 
\graphicspath{{logos/}{figures/}}

\AtlasTitle{Performance of the reconstruction of large impact parameter tracks in the inner detector of ATLAS}
 
\AtlasVersion{2.1}
 
\PreprintIdNumber{CERN-EP-2023-067}
\AtlasJournalRef{Eur. Phys. J. C 83 (2023) 1081}
\AtlasDOI{10.1140/epjc/s10052-023-12024-6}
 
\AtlasAbstract{
Searches for long-lived particles (LLPs) are among the most promising avenues
for discovering physics beyond the Standard Model at the Large Hadron Collider
(LHC).  However, displaced signatures are notoriously difficult to identify
due to their ability to evade standard object reconstruction strategies.  In
particular, the ATLAS track reconstruction applies strict pointing
requirements which limit sensitivity to charged particles originating far from
the primary interaction point.  To recover efficiency for LLPs decaying within
the tracking detector volume, the ATLAS Collaboration employs a dedicated
large-radius tracking (LRT) pass with loosened pointing requirements. During
Run 2 of the LHC, the LRT implementation produced many incorrectly
reconstructed tracks and was therefore only deployed in small subsets of
events. In preparation for LHC Run 3, ATLAS has significantly improved both
standard and large-radius track reconstruction performance, allowing for LRT
to run in all events. This development greatly expands the potential
phase-space of LLP searches and streamlines LLP analysis workflows.  This paper
will highlight the above achievement and report on the readiness of the ATLAS
detector for track-based LLP searches in Run 3.
}
 
\author{The ATLAS Collaboration}

\AtlasRefCode{IDTR-2021-03}

\AtlasJournal{EPJC}

\AtlasRefCode{IDTR-2021-03}
 
\AtlasNote{IDTR-2021-03}

\hypersetup{pdftitle={ATLAS document},pdfauthor={The ATLAS Collaboration}}
 
\begin{document}
 
\maketitle
 
\tableofcontents
 
\clearpage
 
\section{Introduction}
\label{sec:intro}

One promising way to search for physics beyond the Standard Model (BSM) is at
the energy frontier, probing energies comparable to those present very shortly
after the Big Bang.  Thus far, the overwhelming majority of searches for new
physics at CERN's Large Hadron Collider (LHC)~\cite{Evans:2008zzb} have been
performed under the assumption that the BSM particles decay promptly, i.e., very
close to the proton--proton ($pp$) interaction point (IP).  However, BSM theories
often predict the existence of particles with relatively long proper lifetimes
($\tau$)~\cite{HNLtheory, HiddenValey, Alimena_2020}.  For particles moving
close to the speed of light ($c$), this can lead to macroscopic displacements
between the production and decay points of a new long-lived particle (LLP).
When $\tau$ is of $\mathcal{O}(0.1)- \mathcal{O}(10)$~ns, the decay will
predominantly take place within the ATLAS inner detector (ID).  Searches for
LLPs have been performed by the ATLAS Collaboration in the
past~\cite{SUSY-2016-08,SUSY-2017-04,EXOT-2018-61,SUSY-2018-33,SUSY-2018-14,EXOT-2018-57,EXOT-2019-29,SUSYDV_2023},
reconstructing explicitly their decay vertices within the ATLAS ID volume.
 
In order to reconstruct the trajectories of charged particles, ATLAS uses the ID
tracking system to provide efficient, robust and precise position measurements
of charged particles as they traverse the detector.  The energy deposits from
charged particles (hits) recorded in individual detector elements of the ID are
used to reconstruct their trajectories (tracks) and estimate the associated
track parameters.  The standard track reconstruction pass (primary pass) in
ATLAS is optimised for particles that originate at the primary IP, which are
referred to as primary particles. Here, a pass is defined as a single track-finding iteration. In the primary pass, candidate tracks are
required to originate close to the IP thus keeping the computation time
reasonable while still being highly efficient in reconstructing primary
particles.  Several dedicated algorithms are then used to reconstruct a limited set
of specific topologies of displaced decays from well-known Standard Model (SM) particles, such as
from photon conversions and $b$-hadron decays. However, the primary pass does not
efficiently reconstruct charged particles from LLP decays that are displaced by
more than 5\,mm from the IP in the transverse plane, except for specific cases involving photon conversions.  Consequently,
an additional dedicated track reconstruction pass, the so-called
{\textit{large-radius tracking pass}} (LRT pass), is required to extend the
ATLAS sensitivity to a larger particle lifetime range.
 
\begin{figure}[!htb]
\centering
\includegraphics[width=1\linewidth]{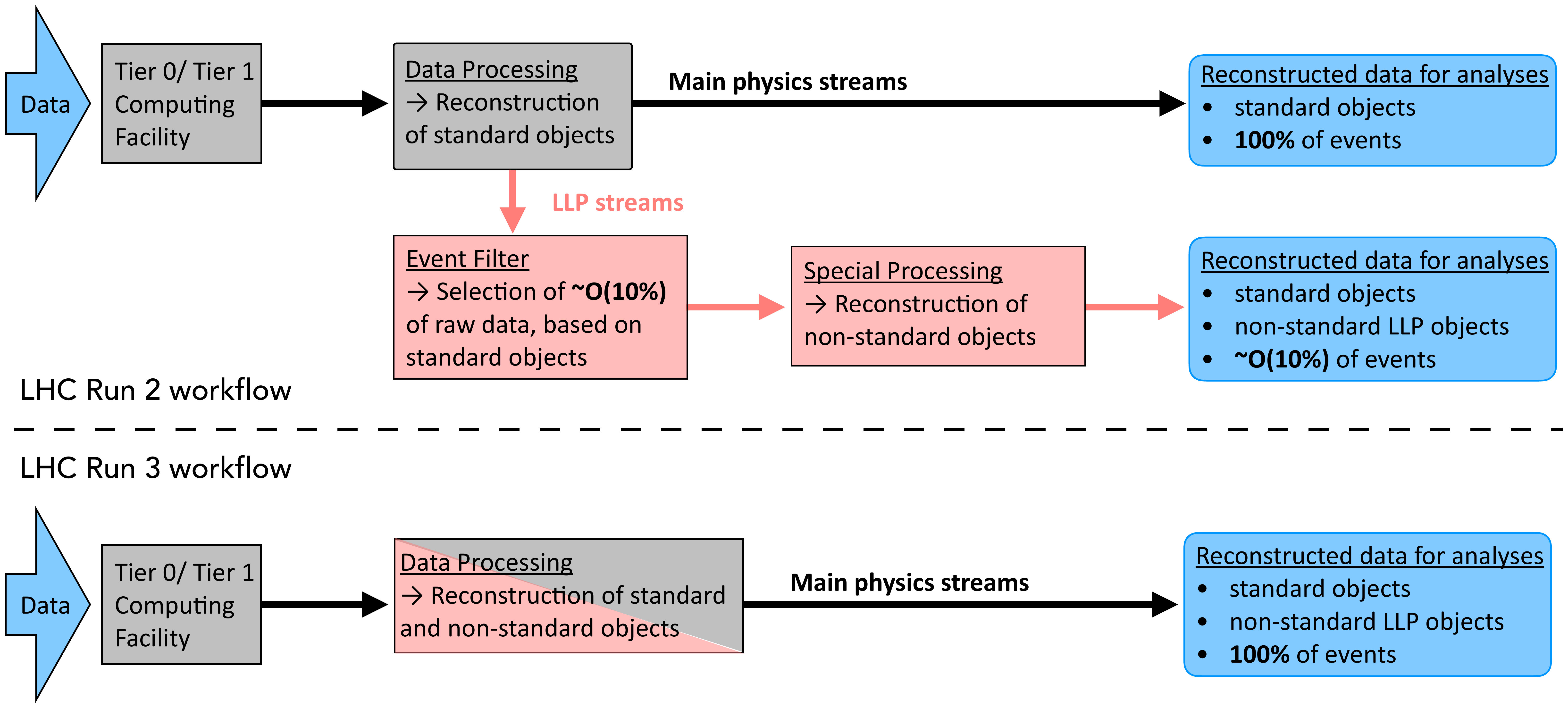}
\caption{Schematic diagram illustrating the LHC Run 2 (top) and improved Run 3
(bottom) event reconstruction workflows for the main physics and LLP streams. ATLAS
defines a hierarchical ``Tier'' structure for computing infrastructure, where
the Tier 0 farm at CERN is used for initial reconstruction of the recorded
data and later data reprocessing proceeds primarily on the eleven Tier 1
sites~\cite{DAPR-2018-01}.}
\label{fig:flow-chart}
\end{figure}

For the LHC Run 2 data-taking period (2015-2018) the LRT pass was solely
optimised for high signal efficiency, resulting in approximately twice the
computational time per event compared to the primary
pass~\cite{ATL-PHYS-PUB-2017-014}. To accommodate this increased processing time,
the LRT pass was only applied to approximately $10\%$ of the events recorded by the ATLAS
detector in a dedicated data stream processed whenever computing resources were
available.  For the LHC Run 3 (2022-2025) data-taking, the average number of
simultaneous $pp$ collisions per bunch crossing, $\langle \mu \rangle$, is
expected to increase to well over 50 from Run 2 values that ranged between 20 and 40.
As $\langle \mu \rangle$ increases, so does the number of potential hit combinations and
therefore the computational time of the LRT pass.  In
preparation for these challenging Run 3 conditions, the LRT pass was updated,
significantly reducing the computation time for the track reconstruction of LLP
signatures and drastically reducing the number of so-called fake tracks due to
random hit combinations.  In parallel, the computation time of the primary pass was improved by a factor of
two~\cite{ATL-PHYS-PUB-2021-012}.  As a consequence of these improvements, a
full integration of the LRT reconstruction for LLP signatures into the standard
ATLAS event reconstruction chain was achieved.  This will enable all ID-based
LLP searches, including new searches using Run 2 data\footnote{A full re-processing of all events recorded by ATLAS during the LHC Run 2 was completed using the updated workflow allowing a seamless combination of datasets in the future.}, to use common data streams, eliminating the need to use a dedicated
processing workflow for this application used in Run~2. A schematic diagram illustrating both event
reconstruction workflows is shown in Figure~\ref{fig:flow-chart}.
 
This paper describes the ``offline'' reconstruction performance of this dedicated
LRT pass. The updated LRT track reconstruction was also
implemented in the ``online''\footnote{Here, ``online'' refers to services and
reconstruction algorithms that run during real-time data-taking at ATLAS.} reconstruction
chain leading to similar performance improvements and consequently new LLP
trigger algorithms for Run 3. The ``online'' performance is out of the scope of
this paper.
The paper is organised as follows: a brief description of the ATLAS detector is
given in Section~\ref{sec:detector}.  Section~\ref{sec:samples} lists the Monte
Carlo (MC) simulation and data samples used to determine the performance.
Section~\ref{sec:tracking} presents the formalism of the ATLAS track
reconstruction and the dedicated LRT pass. Detailed performance results based
both on simulation and data are presented in Sections~\ref{sec:perfSim}
and~\ref{sec:data_mc_compare}, respectively. Concluding remarks are made in
Section~\ref{sec:conclusion}.


\section{The ATLAS detector}
\label{sec:detector}

The ATLAS detector~\cite{PERF-2007-01} at the LHC is a multipurpose particle
detector with a forward-backward symmetric cylindrical geometry that covers
nearly the entire solid angle around the collision point.\footnote{The ATLAS
reference frame is a right-handed Cartesian coordinate system, where the origin
is at the nominal $pp$ interaction point, corresponding to the centre of the
detector. The positive $x$-axis points to the centre of the LHC ring, the
positive $y$-axis points upwards and the $z$-axis points along the beam
direction.
Cylindrical coordinates $(r,\phi)$ are used in the transverse plane, $\phi$
being the azimuthal angle around the beam pipe. The pseudorapidity is defined in
terms of the polar angle $\theta$ as $\eta=-\ln\tan(\theta/2)$. Angular distance
is  measured in units of $\Delta R \equiv  \sqrt{(\Delta\eta)^2 +
(\Delta\phi)^2} $.} It is composed of the ID, electromagnetic and hadronic
calorimeters, and a muon spectrometer.  Lead/liquid-argon sampling calorimeters
provide electromagnetic energy measurements with high granularity and a hadronic
(steel/scintillator-tile) calorimeter covers the central pseudorapidity range up
to $|\eta| < 1.7$. The end-cap and forward regions are instrumented with
liquid-argon calorimeters for both the electromagnetic and hadronic energy
measurements up to $|\eta| = 4.9$. The outer part of the detector consists of a
muon spectrometer with high-precision tracking chambers for coverage up to
$|\eta| = 2.7$, fast detectors for triggering over $|\eta| < 2.4$, and three
large superconducting toroid magnets with eight coils each.  The ATLAS detector
has a two-level trigger system to select events for offline
analysis~\cite{ATL-DAQ-PUB-2016-001}.
An extensive software suite~\cite{ATL-SOFT-PUB-2021-001} is used in data simulation, in the reconstruction and analysis of real and simulated data, in detector operations, and in the trigger and data acquisition systems of the experiment.
 
\subsection{ATLAS inner detector}\label{subsec:ID}
\begin{figure}[ht!]
\centering
\includegraphics[width=0.7\textwidth]{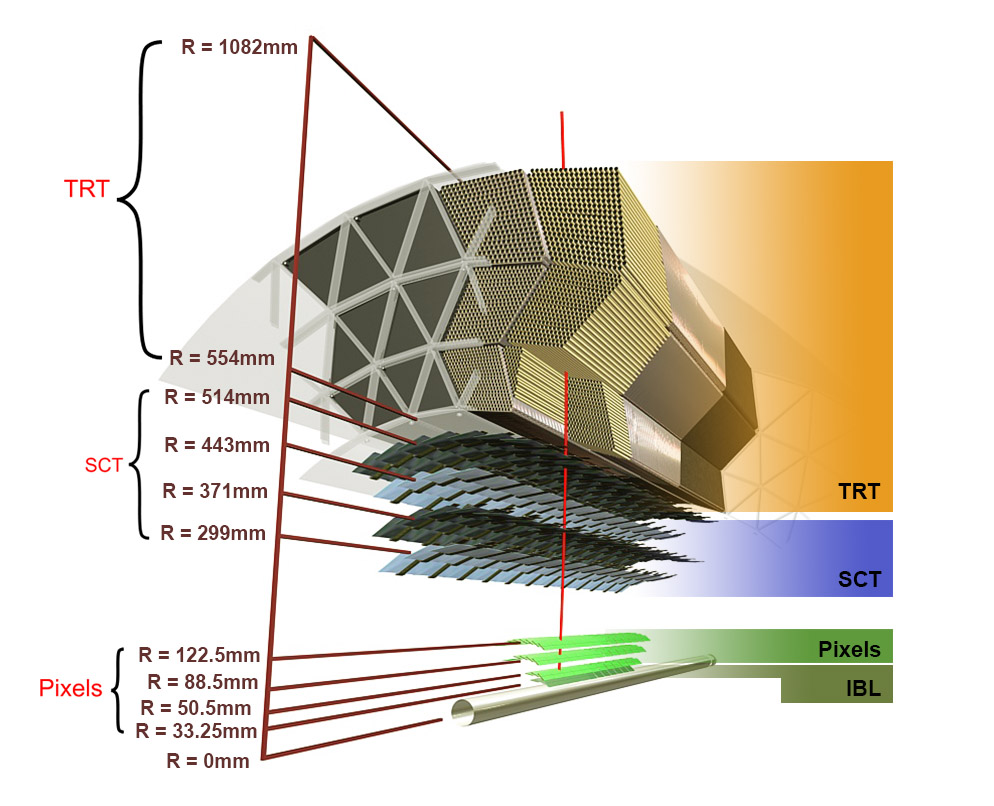}
\caption{A 3D visualisation of the structure of the barrel of the ID. The beam-pipe, the IBL, the Pixel layers, the four cylindrical layers of the SCT and the 72 straw layers of the TRT are shown.}
\label{fig:3Dview}
\end{figure}

The ATLAS ID~\cite{ATLAS-TDR-04,ATLAS-TDR-05} is composed of three subdetectors
utilising three technologies: silicon pixel detectors, silicon strip detectors
and straw drift tubes, all  surrounded by a thin superconducting solenoid
providing a \SI{2}{T} axial magnetic field.  The ID is designed to efficiently reconstruct
charged particles within a pseudorapidity range of $|\eta|<2.5$.
\Cref{fig:3Dview} shows a schematic view of the ID barrel region and a list of the main detector characteristics is provided in \Cref{Tab:IDcharacteristics}.
The material distribution inside the ID has been studied in data through use of
hadronic interactions and photon conversion vertices \cite{PERF-2015-07}. During
the second LHC data taking run with $pp$ collisions at a centre-of-mass energy
$\sqrt{s}=\SI{13}{TeV}$, the ID collected data with an efficiency greater than
\SI{99}{\percent} \cite{DAPR-2018-01}.
 
The innermost part of the ID consists of a high-granularity silicon pixel
detector that includes the insertable B-layer
(IBL)~\cite{ATLAS-TDR-19,PIX-2018-001}, a tracking layer added for Run 2
which is closest to the beam line and designed to improve the precision and
robustness of track reconstruction. The IBL comprises 280 silicon pixel
modules arranged on 14 azimuthal carbon fibre staves surrounding the beam pipe
at an average radius of 33.25~mm. The remainder of the Pixel
detector~\cite{ATLAS-TDR-04,ATLAS-TDR-05} is made of 1744 silicon pixel modules
arranged in three barrel layers and two end caps with three disks each.
In order to simplify the notation throughout the remainder of this paper, the term \textit{Pixel} will be used to refer to
the entire subdetector utilising the silicon pixel detectors, including IBL.
 
The Semiconductor Tracker (SCT)~\cite{IDET-2013-01} contains \num{4088}
silicon strip modules.  They are arranged in four barrel layers and two end caps
with nine disks each.  Each module consists of two pairs of single sided strip
sensors glued back-to-back (axial and stereo sides) with a \SI{40}{mrad} angle.
 
The Transition Radiation Tracker (TRT)~\cite{IDET-2015-01} is the outermost
subdetector of the ID and extends track reconstruction radially outwards to a
radius of 1082~mm.  It is made of \num{350848} gas-filled straw tubes of 4~mm
diameter. The tubes are arranged in 96 barrel modules in 3 layers (32 modules per layer) and 40 disks in each end-cap.
Drift time measurements from the straw tubes are translated into calibrated drift circles.
These allow the TRT to add an average of about 30 two-dimensional additional points to reconstructed tracks that have $|\eta| < 2.0$.
 
The expected hit resolutions and layer positions for each subdetector are
summarised in~\Cref{Tab:IDcharacteristics}. In order to efficiently reconstruct tracks from LLP decays inside the ID, the
radial position of the respective detection layers is crucial. In this context,
LRT has good efficiency up to radius of approximately 300~mm, which corresponds
to the radii of the first SCT barrel layer. Particle decays beyond 300~mm
generally do not create a sufficient number of silicon hits to be reliably reconstructed.

\begin{table}[ht!]
\centering
\caption{Summary of the main characteristics of the ID subdetectors.  The
intrinsic resolution of the Pixel sensors are reported along $r$--$\phi$ and
$z$, while for SCT and TRT only the resolution along $r-\phi$ is given.}
\begin{tabular}{lcl}
\toprule
Subdetector&Intrinsic resolution [\si{\micro\metre}]& Barrel layer radii [\si{mm}]\\
\midrule
Pixel (IBL)&  8 $\times$ 40 &33.2\\
Pixel (non IBL) &10 $\times$ 115&50.5,\,\,88.5,\,\,122.5\\
SCT&17&299,\,\,\,371,\,\,\,443,\,\,\,514\\
TRT&130&from 554 to 1082\\
\bottomrule
\end{tabular}
\label{Tab:IDcharacteristics}
\end{table}


\section{Simulation and data samples}
\label{sec:samples}

To analyze the performance of the LRT pass on BSM physics of interest, three
representative BSM scenarios are studied.  Each of these scenarios are
characterised by the production of long-lived neutral particles which subsequently decay into
charged SM particles.  These signatures have a high probability of producing
charged particles in the ID, and therefore to produce a high number of displaced
tracks.  The choice of the three representative benchmark scenarios is driven by
the different decay chains that characterise the event topology and impact the
kinematics.
 
\begin{figure}[htbp]
\centering
\subfloat[Long-lived neutralino]{\includegraphics[width=.3\textwidth]{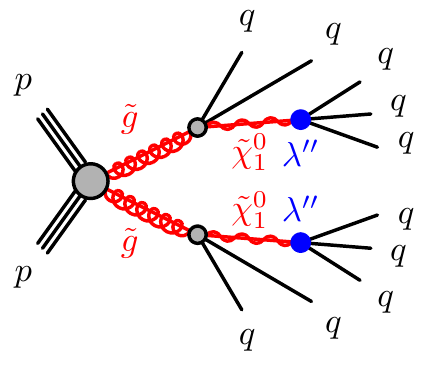}\label{fig:feynman_susy}}\qquad
\subfloat[Higgs portal]{\includegraphics[width=.3\textwidth]{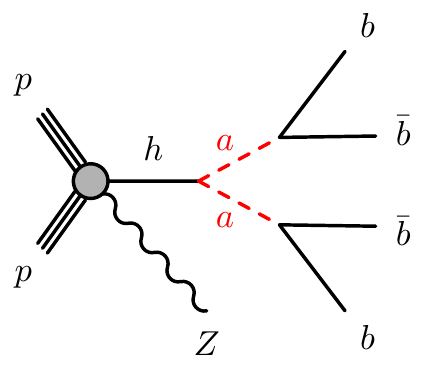}\label{fig:feynman_h4b}}\qquad
\subfloat[Heavy neutral lepton]{\includegraphics[width=.3\textwidth]{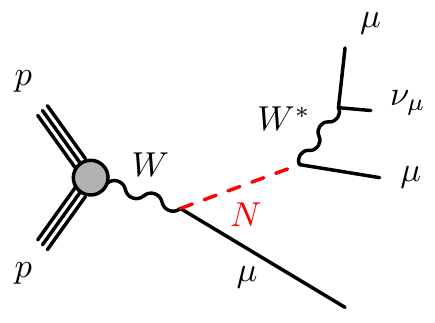}\label{fig:feynman_hnl}}
\caption{Representative Feynman diagrams of models used to characterise the performance of the LRT algorithm.}
\label{fig:1}
\end{figure}

The first signal is a Supersymmetric model where pair production of
gluinos subsequently cascade to a final state with a large number of quarks.
In the sample used, each gluino ($\tilde{g}$) decays into two
quarks and a neutralino ($\tilde{\chi}^0_1$), which, in turn, decays into three
quarks via the R-parity violating coupling
$\lambda^{\prime\prime}$~\cite{Farrar:1978xj,Hall:1983id,Ross:1984yg,Barger:1989rk,Dreiner:1997uz,Barbier:2004ez,Allanach:2003eb}.
Hence, from the gluino pair production the final state at tree level consists of ten
quarks (where $q = u, d, s, c$).  A sketch of the physics process is shown in
Figure~\ref{fig:feynman_susy}.  The mass of the gluino and neutralino are set to
1.6~TeV and 50~GeV respectively and the neutralino coupling is set such that the
proper decay length $c\tau = 300$~mm.  This set of parameters allows to study the performance
of the track reconstruction algorithm on high-multiplicity hadronic tracks with high momentum
originated from the displaced vertex.  The signal samples are generated at
leading-order (LO) accuracy with up to two additional partons in the matrix
element using the \texttt{MadGraph5\_aMC@NLO v2.3.3} event
generator~\cite{Alwall:2014hca} interfaced with \texttt{Pythia
8.212}~\cite{Sjostrand:2014zea}, using the
\texttt{A14}~\cite{ATL-PHYS-PUB-2014-021} tune for the underlying event. The
parton luminosities are provided by the \texttt{NNPDF2.3LO} PDF set~\cite{NNPDF:2014otw}.

The second signal explored is an exotic Higgs boson decay inspired by hidden-sector
models with a Higgs
portal~\cite{Schabinger:2005ei,Patt:2006fw,Strassler:2006im,Strassler:2006ri}.
In this model, a Higgs boson $h$ produced in association with a $Z$ boson decays
into two electrically neutral pseudoscalar particles ($a$) which each decay into a
pair of $b$ quarks, as shown in Figure~\ref{fig:feynman_h4b}.  In the sample
studied, the mean proper decay length and mass of the $a$ boson are
set to 100~mm and 55~GeV respectively, such that a sizeable fraction of displaced vertices formed by the $a$ decays are
produced inside the ID.  The $h \rightarrow aa \rightarrow b\bar{b}b\bar{b}$ decay
forms displaced vertices with fewer charged particles and
lower track momenta than the long-lived neutralino signal. The signal tests the tracking algorithms
in an environment with lower track density. Samples of simulated events
with a Higgs boson produced in association with a leptonically-decaying $Z$
boson are generated using
\texttt{Powheg~Box~v2}~\cite{Alioli:2010xd} 
at next-to-leading order (NLO) with up to one additional jet.  The interface with the
\texttt{GoSam} package~\cite{Cullen:2011ac} provides one-loop amplitude
corrections.  The \texttt{PDF4LHC15nlo} PDF set~\cite{Butterworth:2015oua} is
used.  The decay of the Higgs boson into pseudoscalars and their subsequent decay into
$b$-quarks are simulated with \texttt{Pythia 8.212}. The latter is also used to
simulate the parton showering and hadronisation, as well as underlying event,
with the \texttt{AZNLO CTEQ6L1} set~\cite{STDM-2012-23} of tuned parameters.

The last signal considered is a displaced heavy neutral lepton (HNL)~\cite{Asaka:2005pn,Canetti:2012vf,PhysRevD.29.2539,Bondarenko:2018ptm}, since it can isolate
the behaviour of LRT for
displaced muon tracks.  This signal is motivated by theories in which SM
left-chiral neutrinos mix with a theoretical massive right-chiral counterpart
providing an explanation to the origin of the neutrino mass.  The small mixing
between the HNL ($N$) and the neutrinos result in a long proper lifetime of the
BSM particle.  In this particular signal a $W$ boson decays into a muon and a
neutrino, that mixes with the $N$.  The latter subsequently decays into two
muons and a muon neutrino via an off-shell $W$ or $Z$ boson.  A representation of
one of the possible Feynman diagrams of the signal is shown in
Figure~\ref{fig:feynman_hnl}.  The mean proper decay length and mass of the $N$ are
10~mm and 10~GeV respectively, such that the displaced leptons are produced in the
ID.  Differing from the two aforementioned signals, this signature allows to
study the performance of the tracking algorithms for displaced isolated muon tracks in
low-track multiplicity vertices.  \texttt{Pythia 8.212} is adopted to generate the
events with the \texttt{A14}~\cite{ATL-PHYS-PUB-2014-021} set of tuned
parameters and the \texttt{NNPDF2.3LO} PDF set~\cite{NNPDF:2014otw}.
 
\begin{figure}[htbp]
\centering
\subfloat[]{\includegraphics[width=.45\textwidth]{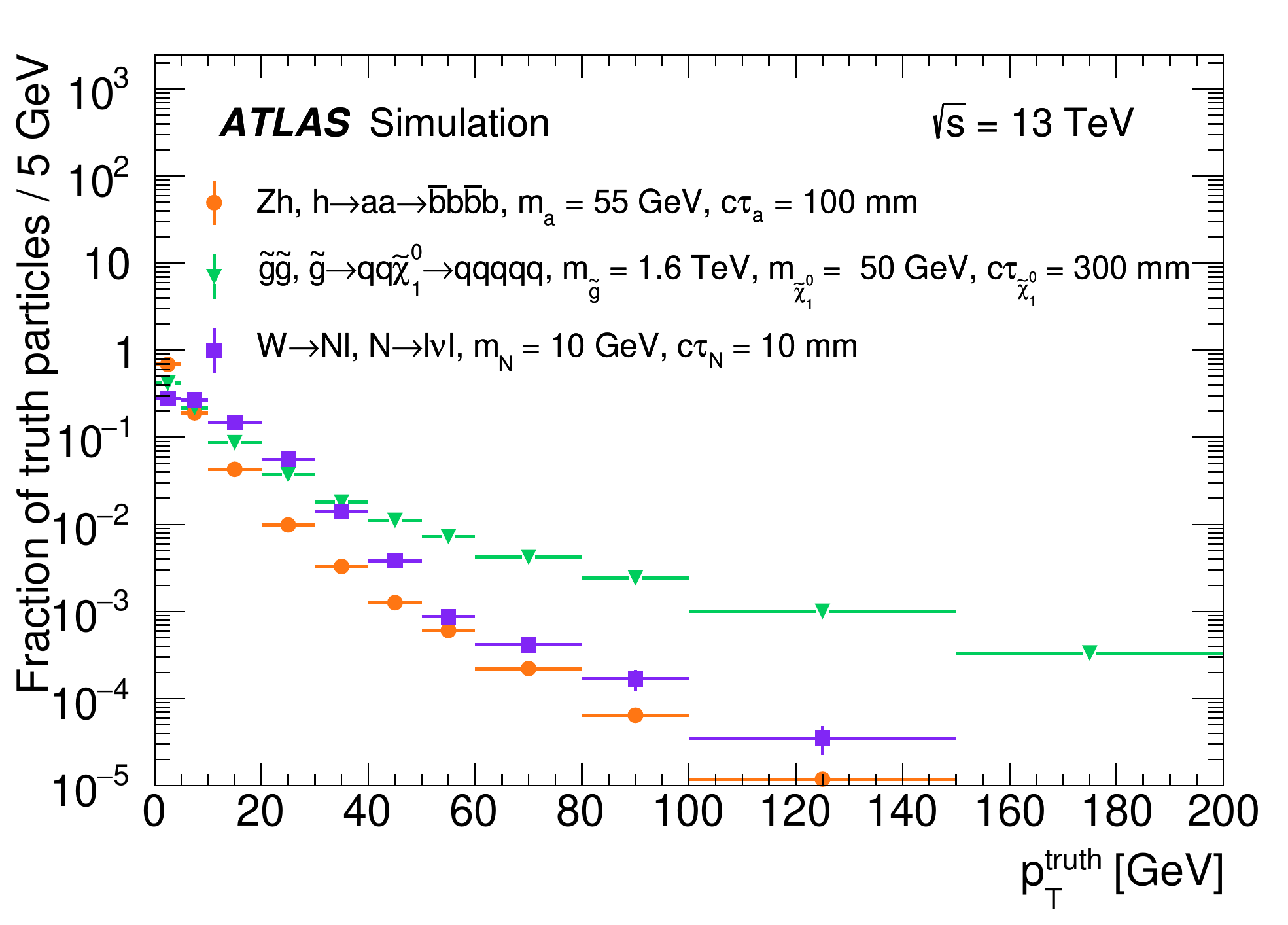}\label{fig:truth_pt}}
\qquad
\subfloat[]{\includegraphics[width=.45\textwidth]{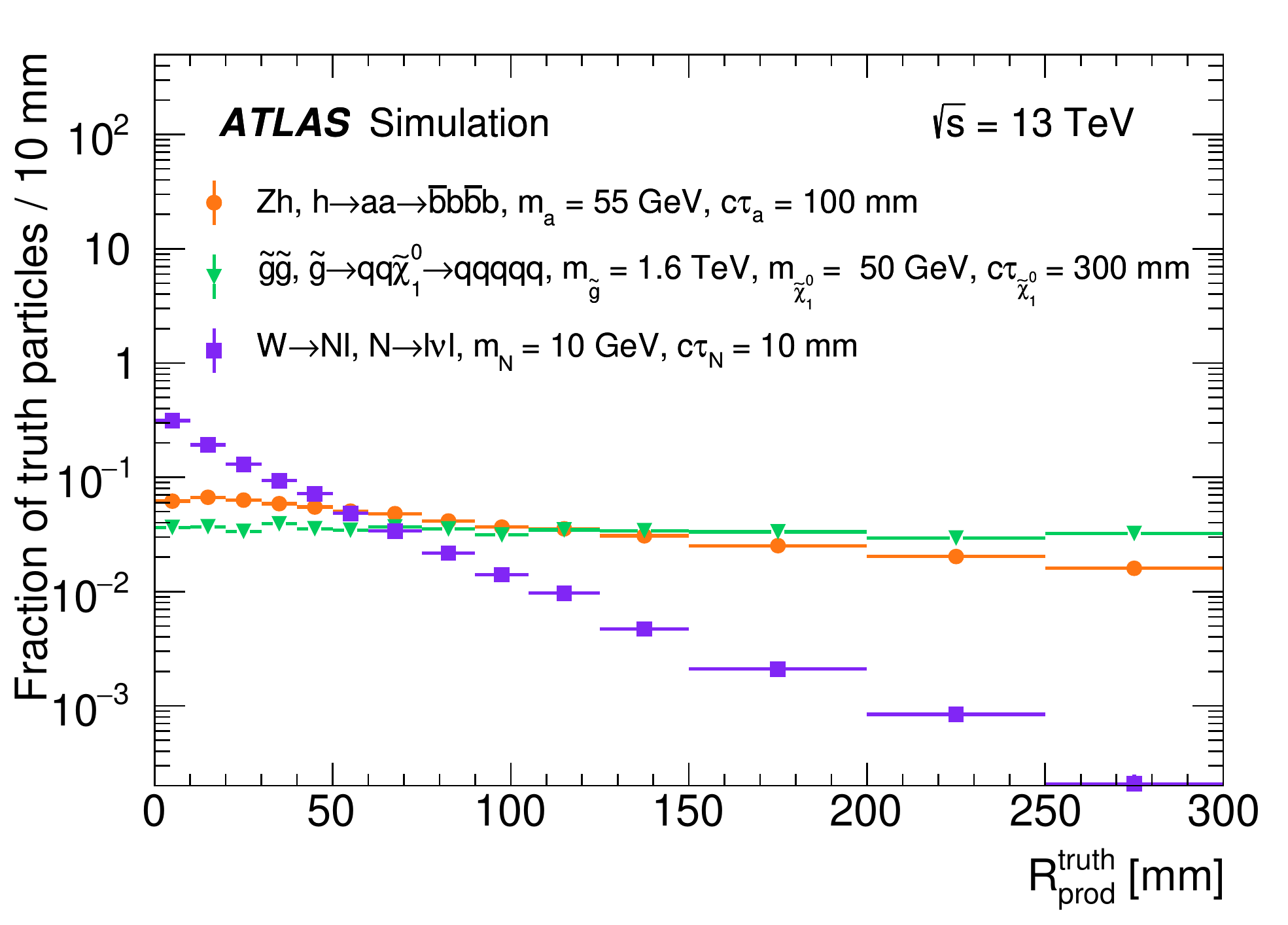}\label{fig:truth_prodR}}\\
\subfloat[]{\includegraphics[width=.45\textwidth]{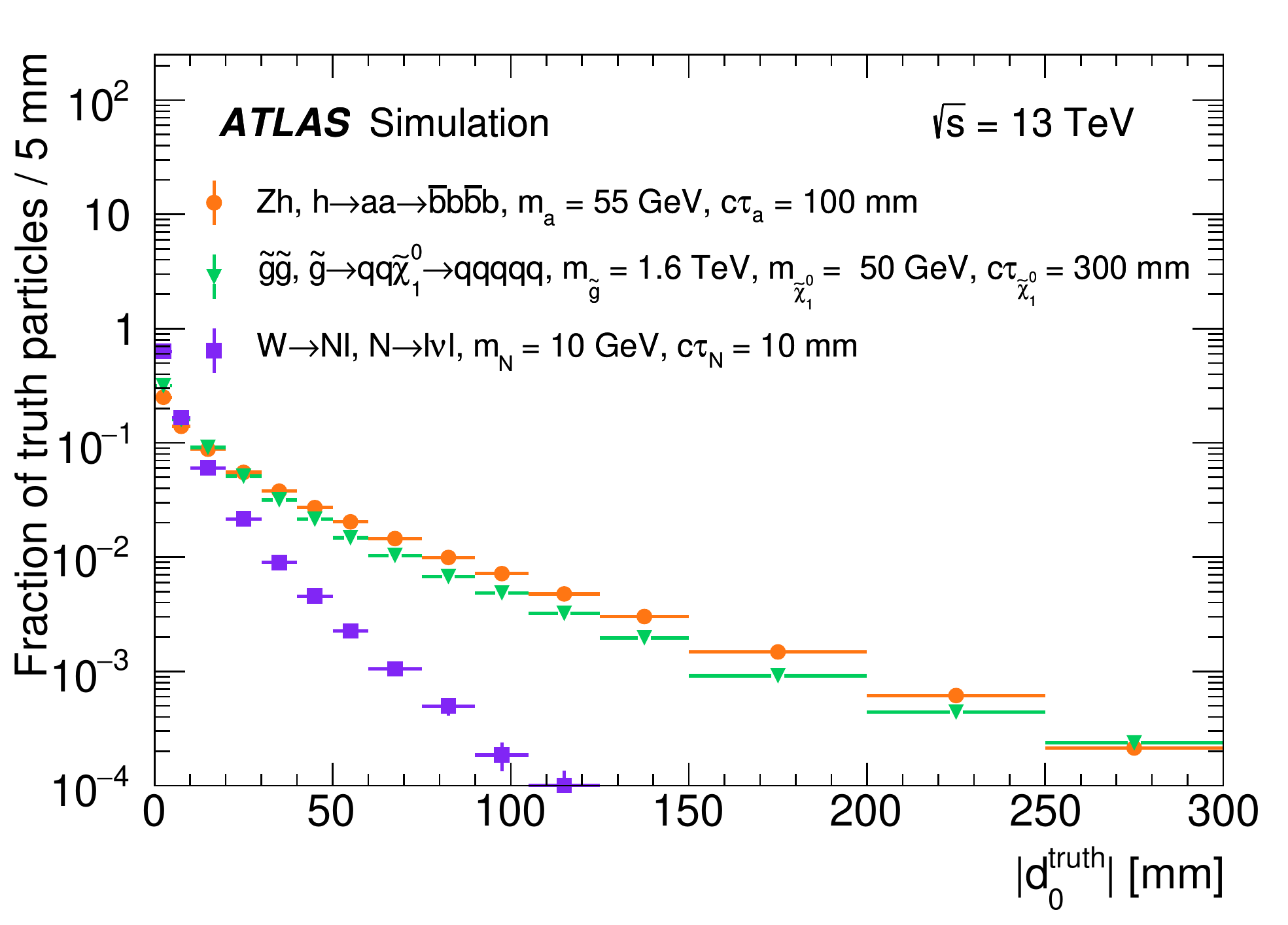}\label{fig:truth_d0}}
\qquad
\subfloat[]{\includegraphics[width=.45\textwidth]{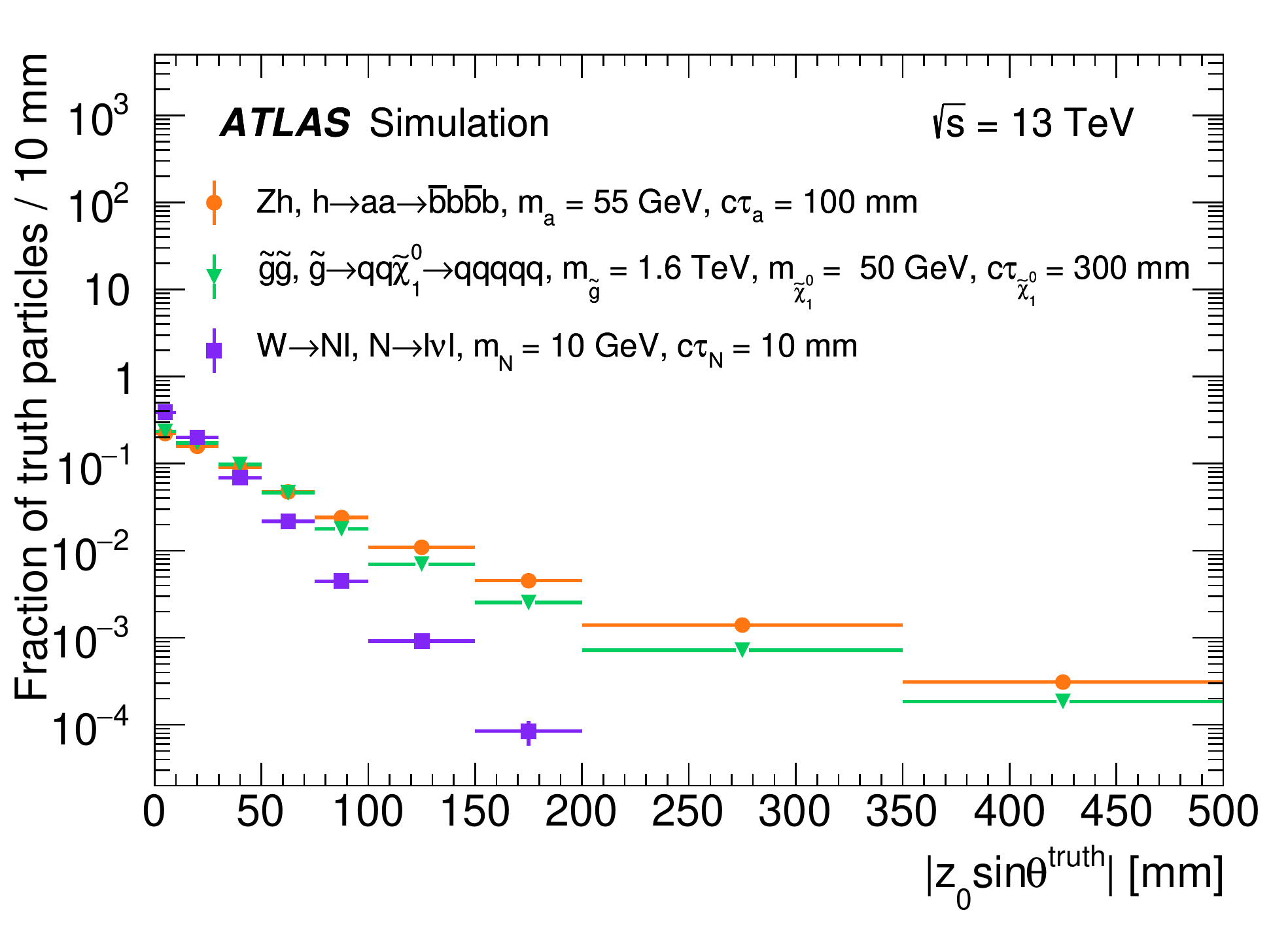}\label{fig:truth_z0sin}}
\caption{Comparison of the (a) $\pt^{\mathrm{truth}}$, (b) $R_{\mathrm{prod}}^{\mathrm{truth}}$, (c)
$|d_{0}^{\mathrm{truth}}|$ and (d) $|z_{0}\sin\theta^{\mathrm{truth}}|$ distributions of the stable charged decay products of the long-lived neutralino, Higgs portal and heavy neutral lepton signal samples.}
\label{fig:truth_kineplots}
\end{figure}

The three signal models vary significantly in terms of the charged particle multiplicity in displaced vertices,
as illustrated in Figure~\ref{fig:1}. In particular, the long-lived neutralino has the highest track multiplicity and
the HNL has the lowest.
Truth particles (particles generated in the MC simulation) are subject to fiducial selection criteria that require the particle to be from a LLP decay within
$R_{\text{prod}} < 400$~mm, and be stable and charged.
Additional selections are imposed on the $\pt>1.2$ GeV and $|\eta|$ < 2.5 of the truth particles.
To further highlight the
different event topologies of the three new physics signatures, a comparison of
the kinematic distributions of charged truth particles is shown in
Figure~\ref{fig:truth_kineplots} as a function the transverse momentum ($\pt^{\mathrm{truth}}$),
the production radius\footnote{The production radius is defined as the radial distance in the transverse plane where the
particle is produced, measured with respect to the detector origin.} ($R_{\mathrm{prod}}^{\mathrm{truth}}$), the transverse impact parameter
($d_{0}^{\mathrm{truth}}$), and the longitudinal impact parameter times $\sin\theta$ ($z_{0}\sin\theta^{\mathrm{truth}}$). All truth quantities use the simulated beam spot as reference.
 
A data sample recorded during 2018 is used to compare the tracking performance
between simulation and data. These are selected from the ``zero bias'' data
stream collected using an unbiased random trigger which fires on the bunch
crossing occurring one LHC revolution after a low-threshold calorimeter-based
trigger~\cite{TRIG-2019-04}.  This ensures that data are collected with equivalent beam conditions to the rest of the standard physics data.
The total number of recorded events is $25,485,193$.  Selected data are compared
to a simulation sample of inelastic $pp$ collisions generated with
\texttt{EPOS}~\cite{Pierog:2013ria} to describe processes with leading jet
$\pt<35$~GeV and with \texttt{Pythia 8.244}~\cite{Sjostrand:2014zea} for events
with a leading jet $\pt>35$~GeV, based on the \texttt{NNPDF2.3LO}
PDF~\cite{NNPDF:2014otw}, using the \texttt{A3}~\cite{ATL-PHYS-PUB-2016-017}
tune.
 
Additional samples are simulated using modified detector geometries to
assess the impact of the modeling of the inner detector material on the track reconstruction
efficiency.  These samples include one with an overall 5\% scaling of the
passive material in ID, one with a 10\% scaling of the IBL, and one with a 25\%
scaling to the pixel services. A fourth variation is simulated using the
QGSP\_BIC physics list instead of the nominal FTFP\_BERT physics
list~\cite{GEANT} to quantify the impact of the choice of parameters in the
GEANT4 simulation.
 
All data and simulation samples were reprocessed using the Run 3 reconstruction
software~\cite{ATL-PHYS-PUB-2021-012}. Simulation samples are generated with a
flat $\langle \mu \rangle$ profile between 0 and 80. In the case of the
comparison with data events, they are reweighted to match the
distribution of $\langle \mu \rangle$ in data with a mean value of 34.


\section{Track reconstruction}
\label{sec:tracking}

 
The primary pass in ATLAS is centred on one
\textit{inside-out} track reconstruction sequence.  This sequence is seeded in
the silicon detectors and additional measurements are added throughout the
ID along the seed trajectory.  In order to improve the reconstruction
efficiency for photon conversion events within the ID, an additional
\textit{outside-in} track reconstruction sequence is performed within so-called
regions of interest (ROI).  The LRT is performed as a third tracking sequence,
following the primary \textit{inside-out} and \textit{outside-in} tracking.
 
\subsection{Primary tracking}\label{sec:std_algo}
 
The primary track reconstruction algorithm first looks for track seeds in the
Pixel and SCT detectors. A seed is a collection of three space points
(SP)\footnote{A SP is defined as a measurement in the Pixel detector or the
combination of axial and stereo layers of the SCT detectors.} each from a unique
layer of the Pixel or SCT detectors, requiring a minimum spatial separation between all
three points. Track seeds are labelled by their middle SP and can only be built from three Pixel or three SCT SPs,
but not combinations of SPs from both detectors. The compatibility of a fourth
SP in a different layer is used to further confirm that the seed is likely to
originate from a charged particle. In this step, a new seed is formed replacing
the outer-most SP of the initial seed and the curvature of the new seed is
compared for compatibility to the original one. A confirmed seed is given high
priority as a promising candidate in the ranked list of seeds.  This list is
composed of all track seeds passing certain quality criteria, including a \pT
and impact parameter selection (see Table~\ref{tab:tracking_details} for details). Here and throughout the rest of the paper, the beam spot~\cite{ATLAS-CONF-2010-027} is used as a reference for all impact parameter calculations.
 
The seeds are used as a starting point to find track candidates using a
combinatorial Kalman filter~\cite{MANKEL1997169,FRUHWIRTH1987444}. This pattern recognition
step combines both measurement association and track fitting in both directions
of the initial seed. Track candidates are then processed by the so-called
ambiguity-solving algorithm~\cite{PERF-2015-08}. The ambiguity solver is
responsible for deciding the final hit assignments in the case of multiple track
candidates using the same hit. Track candidates are evaluated based on various requirements
including track \pt, number of hits, number of shared hits, and
holes\footnote{Hits are defined as measurements from the Pixel, SCT, or TRT
detectors; a hole is defined as a missing hit on an active module where one was
expected based on the particle trajectory; only silicon detectors are considered
in the calculation of holes. Shared hits are those which are used in multiple reconstructed tracks (strong indicator of incorrect
assignments). Modules are the individual silicon sensors.}.
This step removes low-quality
candidates resulting in a collection of high-quality tracks from the track
candidates formed within the pattern recognition stage.
 
Within the ambiguity solver, track candidates are re-fit using a global $\chi^{2}$ method to obtain
the final high-precision track parameter estimate. This method is based on the
Newton–Raphson method and uses an iterative approach to find the best fit to a
set of measurements of a track left in the detector by a charged particle
traversing active detector elements. The quality of the fit is characterised by
track $\chi^{2}$, determined from the distances between the hits in the
detector, which constitutes the track measurements, and the fitted track. The track fit accounts for the effects of multiple Coulomb
scattering of the particle with the detector components. A detailed description
can be found in Ref.~\cite{IDTR-2019-05}. An extension of the track into the
TRT subdetector is attempted, with a re-fit of the entire track being performed
in case of a successful extension to profit from the additional measurements.
 
The \textit{outside-in} sequence is performed using the detector hits not
already assigned to tracks from the \textit{inside-out} reconstruction. The ROIs
are determined by deposits in the electromagnetic calorimeter
with transverse energy $E_{\mathrm{T}}>6$~GeV. This second sequence uses segments of hits in the TRT
that are spatially compatible with the ROI to find 2-SP SCT track seeds, that are extended
inwards into track candidates using the same procedure as for the primary pass
but with reduced hit criteria. A dedicated ambiguity resolution step is performed among
all track candidates in the second pass and a re-fit of the resulting tracks
including their TRT extension completes the \textit{outside-in} sequence.
 
The resulting tracks, from both \textit{inside-out} and \textit{outside-in}
sequences, form the final collection of reconstructed primary tracks.  The
trajectory of these tracks is parameterised by a set of five parameters: $d_0$,
$z_0$, $\phi_0$, $\theta_0$, $q/p$, where $d_0$ and $z_0$ are the transverse and
longitudinal impact parameters and $\phi_0$ and $\theta_0$ the azimuthal and
polar angles of the track, all defined at the point of closest approach to the
$z$-axis of the reference frame~\cite{IDTR-2019-05}. The ratio $q/p$ is the inverse
of the particle momentum ($p$) multiplied by its charge ($q$).
 
\subsection{Large-radius tracking}
\label{sec:lrt_algo}
 
The LRT sequence is executed after the primary tracking iteration using
exclusively the detector hits not already assigned to primary tracks, ensuring a
completely independent collection of tracks. It adopts the same logic as the
primary \textit{inside-out} tracking iteration but it is specifically optimised
for LLP signatures. As mentioned previously, a legacy implementation of LRT was used during the LHC Run 2
data-taking that was optimised for high LLP signal-track reconstruction efficiency at the
expense of a very large fake track rate (see Section~\ref{sec:fakes} for
a detailed definition).
These choices resulted in a computationally-expensive algorithm that could only
be operated on a small fraction of events recorded by the ATLAS experiment in a
dedicated stream.
In preparation for LHC Run 3 the LRT sequence was reoptimised
to substantially reduce the required computational
resources per event, and consequently integrate LRT as a third common tracking
sequence that is executed for each event within the updated default ATLAS event
reconstruction chain.
 
This section gives a detailed description of the reoptimised LRT sequence
and the implemented changes. Two guiding principles were followed during the
optimisation to substantially speed up the processing time and reduce disk usage
per event: 1) identify and terminate the track reconstruction for fake candidate
tracks as early as possible within the reconstruction sequence, and 2) significantly
reduce the overall number of falsely reconstructed tracks while maintaining high
signal track reconstruction efficiency. Table~\ref{tab:tracking_details} lists the most
important configuration differences between the primary tracking and LRT
sequences.
 
\begin{table}[htbp]
\begin{center}
\caption{Most important selection criteria that differ between the primary tracking (\textit{inside-out} sequence only)
and LRT setups. Common selection criteria that were changed from the legacy
implementation are also shown.}

\makebox[0pt]{
\begin{tabular}{l@{\hskip .5in}cc}
\toprule
Selection criteria & Primary (\textit{inside-out})& LRT \\
\midrule
max. $|d_0|$ [mm]& 5 & 300 \\
max. $|z_0|$ [mm]& 200 & 500\\
min. $\pt$ [GeV]& 0.5  & 1 \\
max. $|\eta|$ &  2.7 & 3.0 \\
max. silicon holes & 2 & 1 \\
max. double holes & 1 & 0 \\
max. holes gap & 2 & 1 \\
road width [mm] & 12 & 5 \\
seeding & Pixels and SCT & SCT only\\
max. seeds per middle Pixel SP\qquad & 1 & -- \\
max. seeds per middle SCT SP\qquad& 5 & 1\\
\midrule
\multicolumn{3}{c}{Common selection criteria} \\
\midrule
min. silicon hits & \multicolumn{2}{c}{8} \\
min. unshared silicon hits & \multicolumn{2}{c}{6} \\
max. track $\chi^{2}/n_{\mathrm{DoF}}$ & \multicolumn{2}{c}{9} \\
keep all confirmed seeds &  \multicolumn{2}{c}{true} \\
 
\bottomrule
\end{tabular}
}


\label{tab:tracking_details}
\end{center}
\end{table}
 
LRT tracks are seeded from SCT SPs only and a maximum of one seed per middle SP
is allowed. This configuration is motivated by the displaced topology of tracks
originating from LLP decays that often do not result in at least three Pixel
SPs. No limit on the number of seeds is applied for confirmed seeds (only introduced for the updated LRT) and all
such seeds are kept. To reflect the extended phase space of LRT tracks, criteria
on the estimated impact parameters $d_0$ and $z_0$ are significantly relaxed. In
contrast, the minimum \pt threshold is increased to 1\,GeV to reduce the large
number of soft charged pion tracks created in material interactions. The seed
ordering in the primary pass uses $|d_0|$ as the defining criteria to order the
relevance of seeds for the track finding stage (the seed with the lowest $|d_0|$
is ranked highest). This criteria is still relevant for tracks from LLP decays
as the most probable $|d_0|$ value of such seeds is still small, owing to the
exponential distribution of LLP lifetime. In addition, the LRT seed ordering includes a
$\Delta\eta = |\eta_{\mathrm{seed}}-\eta'|$ weight factor, where $\eta'$ is the
pseudorapidity of the vector connecting the beam spot position with the point of
closest approach of the seed. Seeds from LLP decays tend toward small values of
$\Delta\eta$, due to the correlated direction of the LLP and its decay products, in
contrast to seeds corresponding to fake tracks.  These optimisations
significantly improve the execution time by reducing the number of combinatorial
permutations to process during the seed and track finding stages, without
significantly changing the number of signal tracks being reconstructed.
 
The same relaxed impact parameter selections are imposed in the track finding
and ambiguity solving steps. Other track selection criteria are significantly
stricter for LRT tracks to ensure that a sufficiently low fake track rate is
achieved. Tracks are required to pass the same strict silicon hit and track
$\chi^2/n_{\mathrm{DoF}}$ requirements as primary tracks. In addition, the maximum number of
silicon holes is reduced to one, no double hole\footnote{A double whole is defined as two consecutive missing hits on active modules where both were expected based on the particle trajectory. In this context, the holes gap refers to the largest gap of consecutive missing hits on active modules (used in Table~\ref{tab:tracking_details}).} on a track is allowed, and the
search roads (sets of silicon detector modules that can be expected to contain hits compatible with the seed) used to extend the seeds within the track finding stage are
narrowed. These strict selections reduce the acceptance of tracks from LLP decays by less than
5\% but they significantly reduce the number of low-quality tracks written to
storage. Thus, they also lower the required number of iterations of the ambiguity resolution
and TRT extension phases.
 
\begin{figure}[!t]
\centering
\subfloat{\includegraphics[width=.47\textwidth]{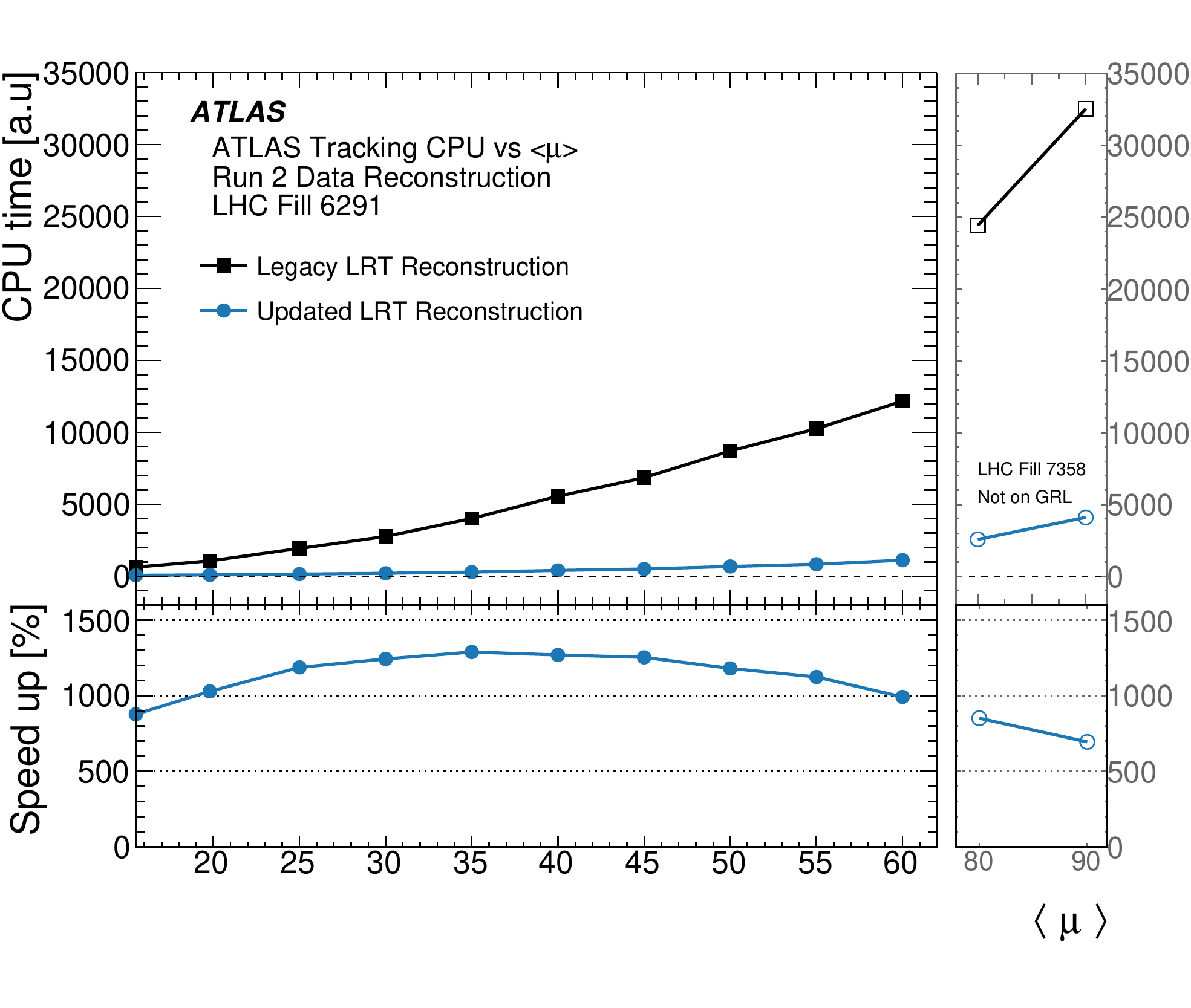}}
\qquad
\subfloat{\includegraphics[width=.47\textwidth]{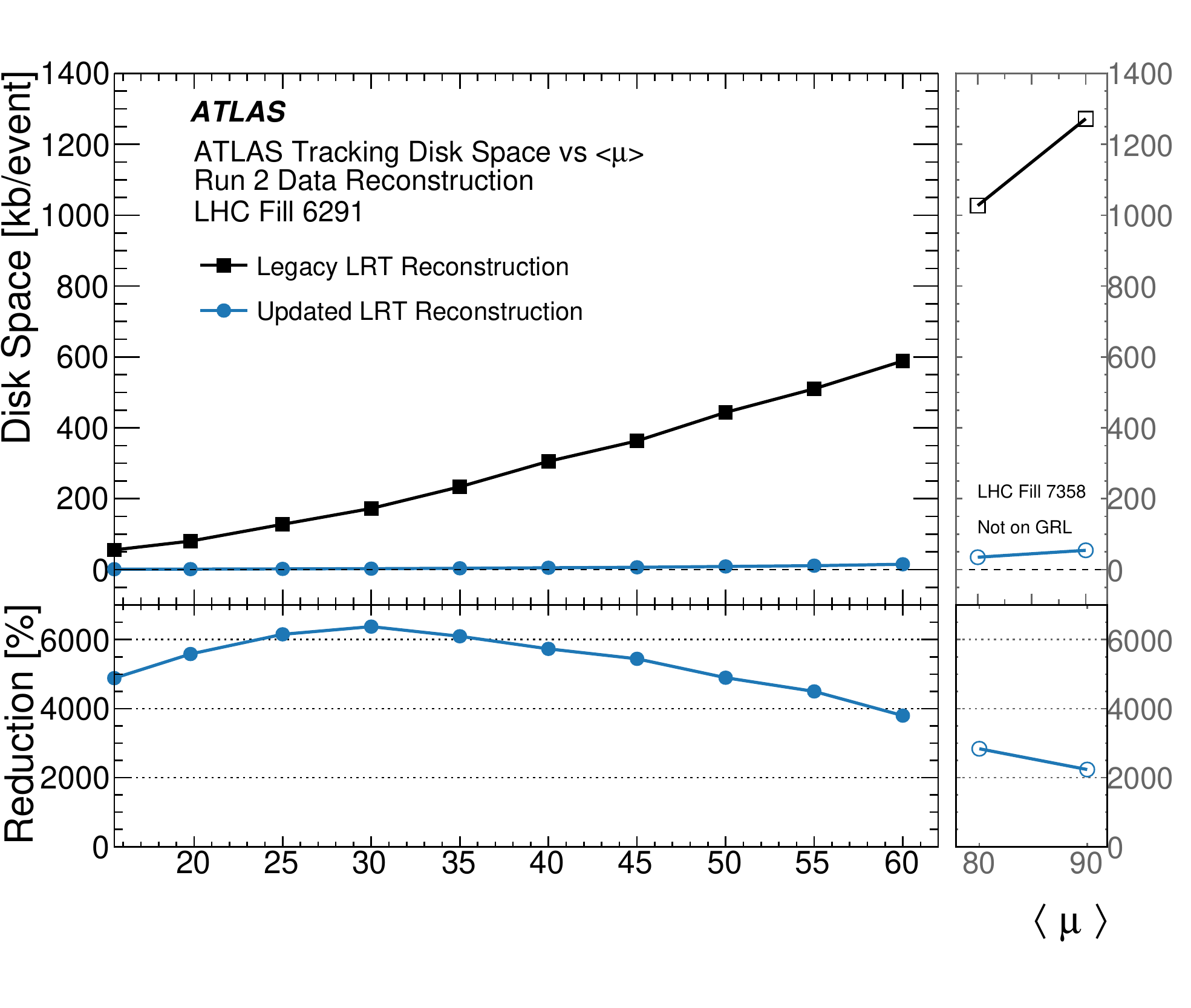}}
\caption{The (a) processing time taken per event to reconstruct the same data events
and (b) event size of the LRT reconstruction output, versus average
pileup, comparing the legacy LRT and updated LRT reconstruction
software. The high average pileup regime (open markers) with $\langle \mu
\rangle >80$ indicates data events taken from a 2018 machine development fill
not passing the full ATLAS data quality requirements and therefore these are not included in the so-called  `Good Run List', GRL (`Not on GRL'). This range is included
to provide an indication of the expected performance in such extreme pileup
regimes.}
\label{fig:resources}
\end{figure}

The resulting collection of tracks, whether they have a valid TRT extension or
not, form the LRT track collection. The reduced total acceptance of LLP signal tracks
compared to the legacy implementation is between 5\% - 10\% independent of the
decay radius of the LLP. Detailed reconstruction efficiency studies of this new
LRT pass are discussed in Sections~\ref{sec:perfSim}
and~\ref{sec:data_mc_compare}. When performing measurements or searches of new phenomena
using the ATLAS data, the Poisson statistics of a counting experiment often results in a sensitivity related to
$N_{\mathrm{s}}/\sqrt{N_{\mathrm{b}}}$, where
$N_{\mathrm{s}}$ corresponds to the signal yield and
$N_{\mathrm{b}}$ refers to the number of background events. A similar
figure of merit, defined with the number of tracks originated from the LLP decays ($N_{\mathrm{trk}}^{\mathrm{LLP}}$) and from fake tracks ($N_{\mathrm{trk}}^{\mathrm{fake}}$),
$N_{\mathrm{trk}}^{\mathrm{LLP}}/\sqrt{N_{\mathrm{trk}}^{\mathrm{fake}}}$ is
used to quantify the impact of LRT on physics analyses.
This figure of merit has improved by more than 400\% for LHC Run 2 pileup conditions, making this third tracking
pass a highly efficient reconstruction sequence.
$N_{\mathrm{trk}}^{\mathrm{LLP}}$ ($N_{\mathrm{trk}}^{\mathrm{fake}}$) is
strongly correlated with $N_{\mathrm{s}}$ ($N_{\mathrm{b}}$), however, this figure of merit here describes the LRT improvement and does not directly translate to equal sensitivity gains for LLP searches.
Figure~\ref{fig:resources} shows the time to process\footnote{All benchmark studies described in this work were run as the only active user on a dedicated machine equipped with an AMD EPYC$^{\mathrm{TM}}$ 7302 16-core processor, running the CERN CENTOS 7 operating system. The processor was operated in `performance' mode, with simultaneous multi-threading (SMT) and frequency boosting disabled.} the LRT reconstruction and the output size of each event as a function of the average interactions per bunch crossing in 2018 data sample (no trigger or event topology selections applied).
The processing time of the new implementation has improved by more than 10 times compared to the
legacy implementation and the disk space usage per event for LRT tracks has been
reduced by more than a factor of 50.
 
\subsection{Truth-matching in simulation and fake tracks}
\label{sec:fakes}
In order to match a generated particle in simulation to a reconstructed track, a
requirement ($R_{\mathrm{match}}$) is applied that is defined by comparing the
number of measurements in detector elements which are common between the truth
particle trajectory and the reconstructed track:
 
\begin{linenomath}
\begin{equation}
R_{\mathrm{match}} = \frac{10 \times N^{\mathrm{Pixel}}_{\mathrm{common}} + 5 \times N^{\mathrm{SCT}}_{\mathrm{common}} +  N^{\mathrm{TRT}}_{\mathrm{common}}}{10 \times N^{\mathrm{Pixel}}_{\mathrm{reco}} + 5 \times N^{\mathrm{SCT}}_{\mathrm{reco}} + N^{\mathrm{TRT}}_{\mathrm{reco}}},
\label{eq:truthmatch}
\end{equation}
\end{linenomath}
 
where $N^{\mathrm{Pixel/SCT/TRT}}_{\mathrm{reco}}$ and
$N^{\mathrm{Pixel/SCT/TRT}}_{\mathrm{common}}$ are the numbers of hits in the
different detectors which are present on the reconstructed particle trajectory,
and present on both the truth particle trajectory and the reconstructed track,
respectively. The different weights applied to the different sub-detector are
motivated by the different number of hits available in each sub-detector for
the track reconstruction and the resolution of those hits. Sub-detectors with
fewer hits are assigned with larger weights to ensure the truth matching does
not favor one particular sub-detector. Reconstructed tracks with high values of
$R_{\mathrm{match}}$ are considered well-matched to a truth particle and are
used in efficiency calculations. Conversely, poorly-matched tracks (low values
of $R_{\mathrm{match}}$) are referred to as \emph{fake tracks}. These tracks are
dominated by incorrect combinations of hits from multiple particles in the
pattern recognition. In the following, $R_{\mathrm{match}} > 0.5$ is used to
identify well-matched tracks, while all other tracks are labelled fake tracks.
 
\subsection{Secondary vertex reconstruction}
\label{sec:vertexing}
 
Many analyses which used LRT in Run 2 attempted to reconstruct the decays of
LLPs by performing an additional secondary vertex reconstruction
using both primary and LRT tracks. To study how the improved LRT reconstruction
performs in these downstream algorithms, two different secondary vertex
reconstruction algorithms are used.
 
The first is a dedicated inclusive vertexing algorithm optimised for identifying
the decays of heavy LLPs ~\cite{ATL-PHYS-PUB-2019-013}. Vertices
are formed from two-track seeds, which are required to contain at least
one track with $|d_0| > 2$~mm. The compatibility of each possible pair of
preselected tracks is then computed, and those deemed loosely-compatible are
retained. These two-track seed vertices are then combined to form multi-track
vertices using a pairwise compatibility graph. Nearby vertices are then merged,
and lower-quality tracks not initially preselected for vertex seeding are
attached to compatible vertices.
 
The second is an algorithm optimised for identifying two-body $V0$
decays (i.e. $K^0$ and $\Lambda^0$)~\cite{STDM-2010-09}. Tracks are considered for vertex reconstruction if
they are not associated to any primary $pp$ interaction vertex (PV)~\cite{ATL-PHYS-PUB-2019-015}, have
$\pt > 1$~GeV, and satisfy $|d_0/\sigma_{d_0}| > 2$, where $\sigma_{d_0}$ is the
uncertainty on the reconstructed $d_0$.  All opposite-charge two-track pairs are
then considered. Vertices are kept if the $\chi^2/n_{\mathrm{DoF}}$ probability
of the two-track fit is greater than 0.0001. To reject vertices from random
track combinations, selections are placed on the transverse ($a_{0,xy}$) and
longitudinal ($a_{0,z}$) point of closest approach between the vertex flight
direction\footnote{The vertex flight direction is defined as the direction of the vertex momentum vector, computed from
the vectorial sum of the individual track momenta vectors.} and the PV with the largest $\sum \pt^2$ of associated tracks. Vertices are kept if they satisfy $|a_{0,xy}|
< 3$~mm and $|a_{0,z}| < 15$~mm. To reject residual contributions from
combinations of prompt tracks, vertices are additionally required to have
$L_{xy}/\sigma_{L_{xy}} > 2$ , where $L_{xy}$ is the transverse displacement of
the vertex with respect to the primary vertex, and $\sigma_{L_{xy}}$ is its
uncertainty.


\section{Performance study on simulation}
\label{sec:perfSim}


\subsection{Reconstruction efficiency}
 
To assess the track reconstruction efficiency in simulation, the hit-based
matching scheme described in Section~\ref{sec:fakes} is used to associate
reconstructed tracks to generated particles.  The track reconstruction
efficiency is then defined as the ratio of the number of signal truth particles
matched to reconstructed tracks divided by the number of signal truth particles.
Truth particles are subject to fiducial selection described in Section~\ref{sec:samples}.
These requirements are chosen to be significantly far away from the selection criteria imposed in the LRT reconstruction in order to avoid turn-on effects due to reconstruction resolutions of track parameters.
 
\begin{figure}[!htb]
\centering
\subfloat[]{\includegraphics[width=.45\textwidth]{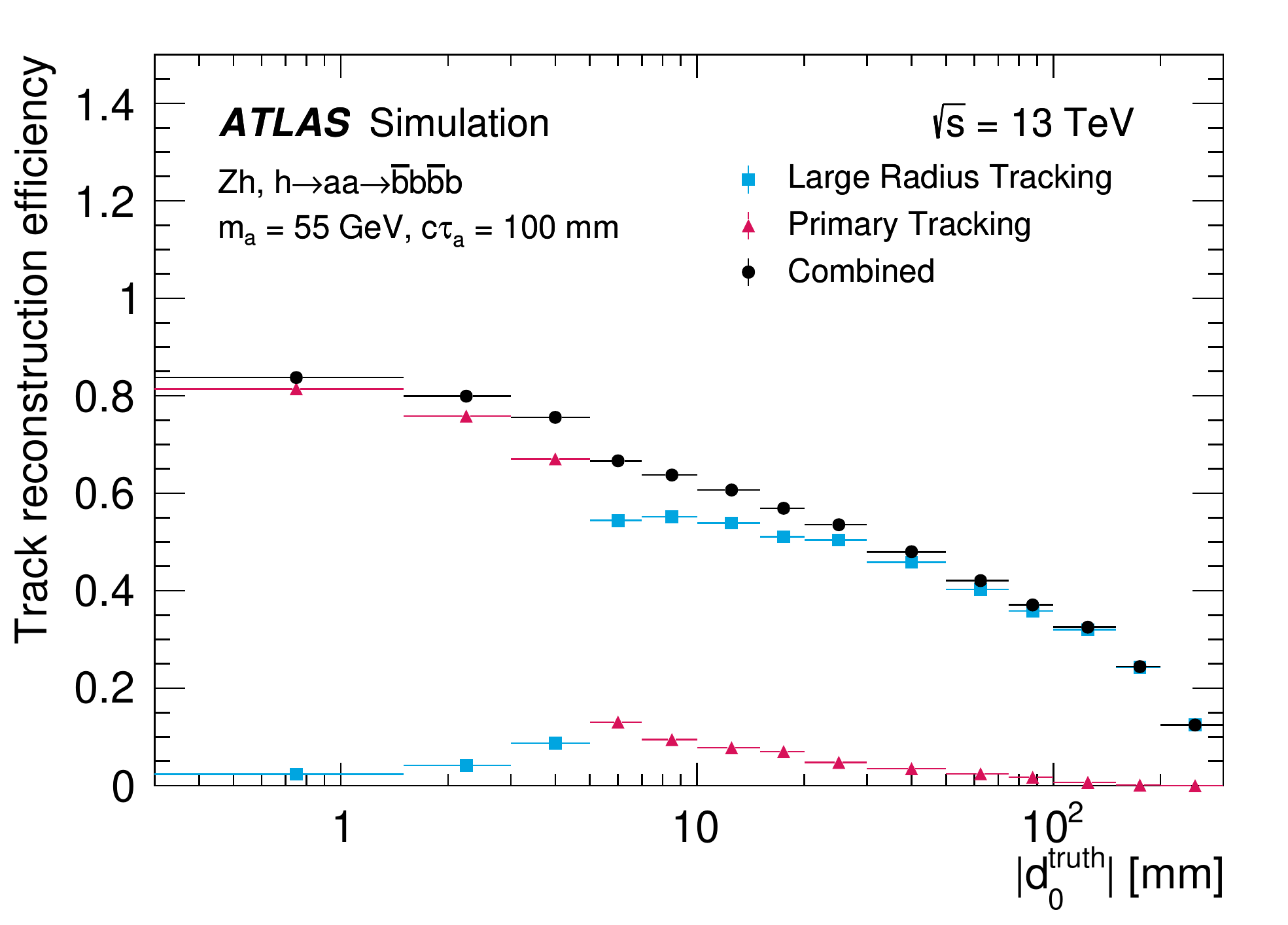}\label{fig:eff_vh4b_d0}}
\qquad
\subfloat[]{\includegraphics[width=.45\textwidth]{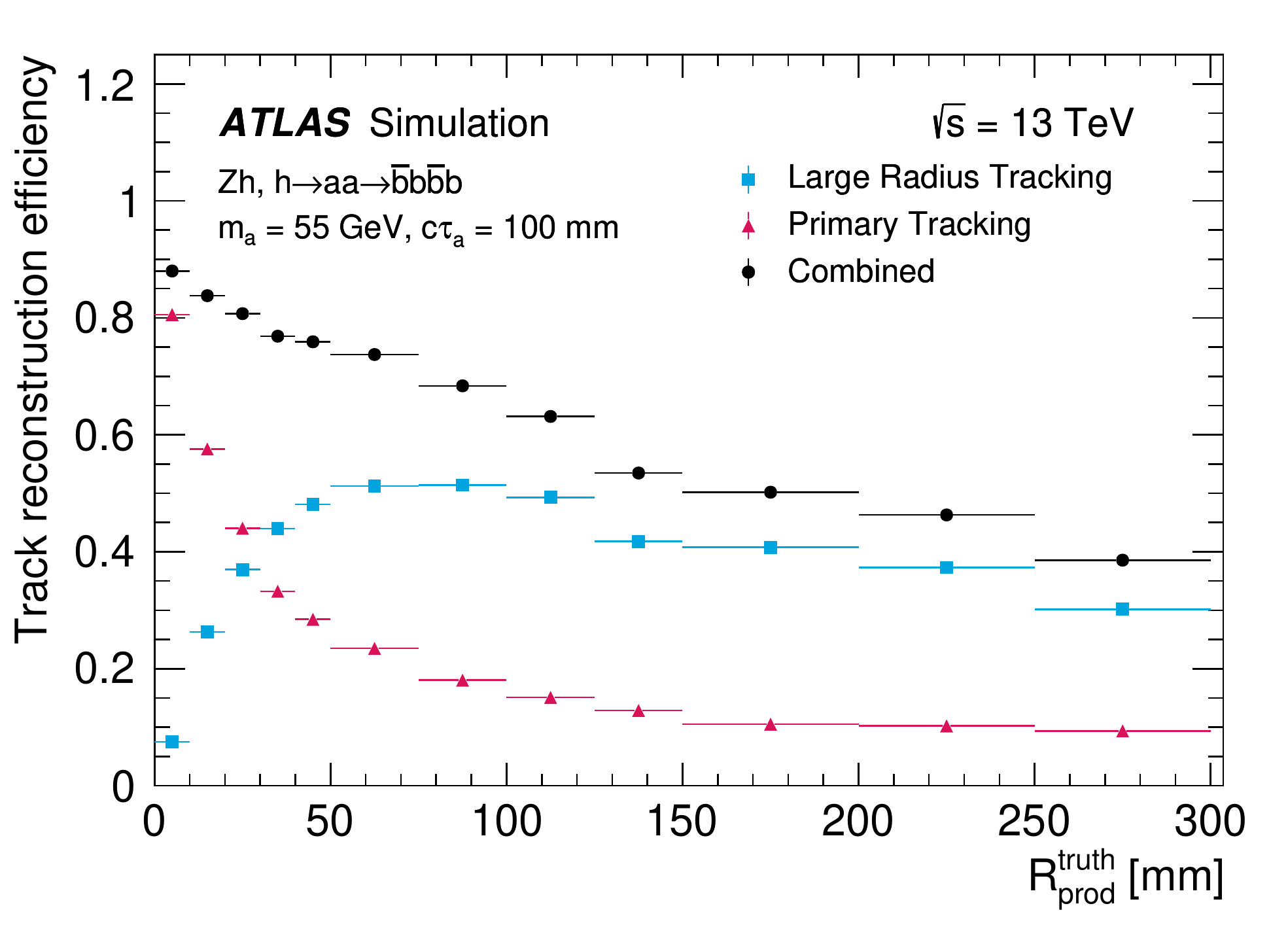}\label{fig:eff_vh4b_r}}
\caption{The primary, LRT, and combined track reconstruction
efficiencies for displaced charged particles produced by the decay of the LLP
in the Higgs portal signal model. Efficiencies are shown as a function of (a)
$|d_{0}^{\mathrm{truth}}|$ and (b) $R_{\mathrm{prod}}^{\mathrm{truth}}$.
The first bin of (a) includes tracks with $|d_{0}^{\mathrm{truth}}|$ down to $|d_{0}^{\mathrm{truth}}| = 0$~mm.
Truth particles are required to fulfill the fiducial selection criteria stated in the text.
}
\label{fig:eff_vh4b}
\end{figure}

\begin{figure}[!htb]
\centering
\subfloat[]{\includegraphics[width=.45\textwidth]{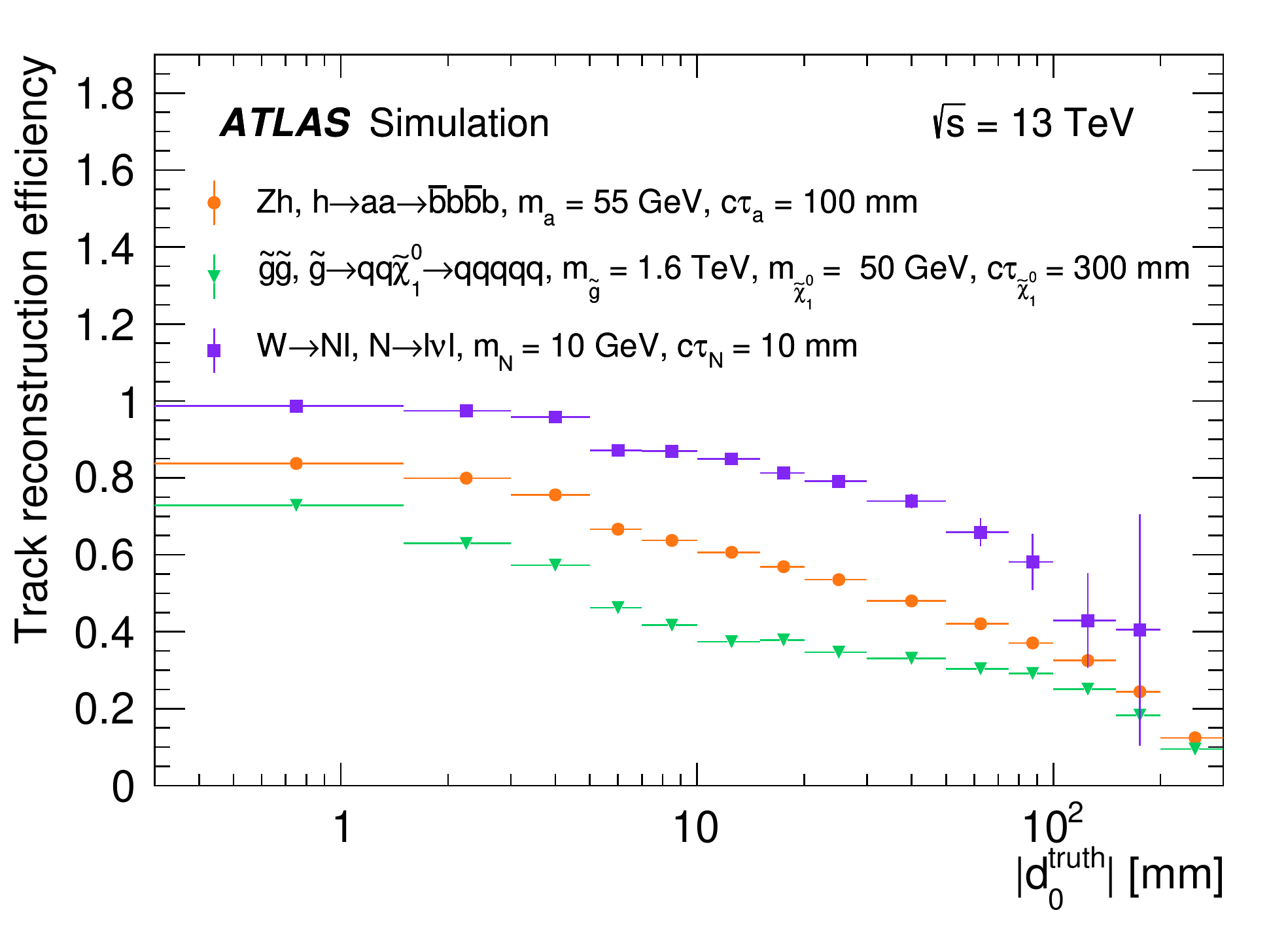}\label{fig:tot_llp_d0}}
\qquad
\subfloat[]{\includegraphics[width=.45\textwidth]{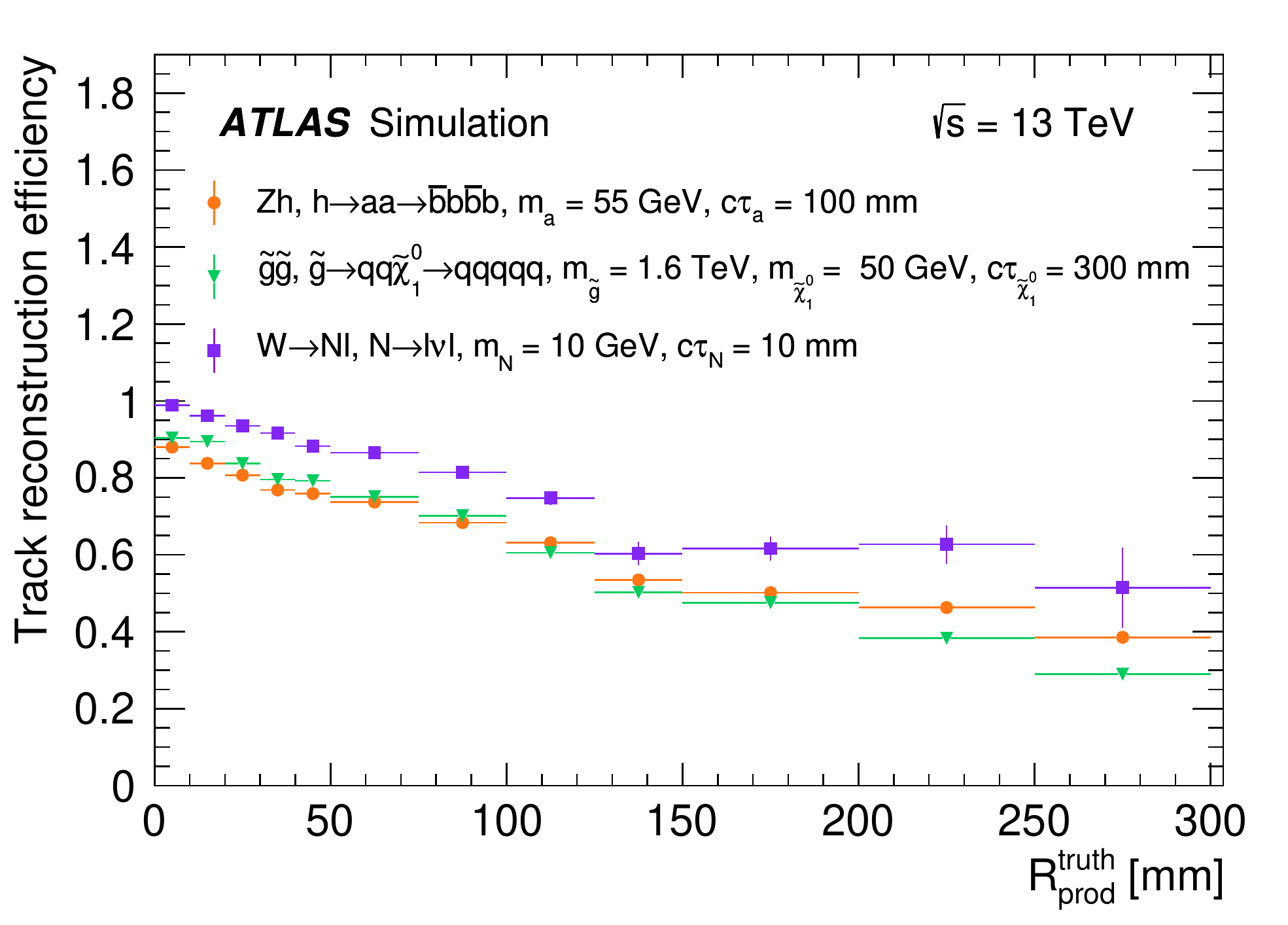}\label{fig:eff_tot_llp_r}}
\caption{Combined track reconstruction efficiency for displaced charged
particles produced by the decay of long-lived signal particles in the
long-lived neutralino, Higgs portal and heavy neutral lepton signal models.
The efficiency is shown as a function of (a) $|d_0^{\mathrm{truth}}|$ and (b)
$R_{\mathrm{prod}}^{\mathrm{truth}}$.
The first bin of (a) includes tracks with $|d_{0}^{\mathrm{truth}}|$ down to $|d_{0}^{\mathrm{truth}}| = 0$~mm.
Truth particles are required to fulfill the fiducial selection criteria stated in the text.
}
\label{fig:eff_tot_llp}
\end{figure}

The track reconstruction efficiencies for the Higgs portal benchmark models are shown in
Figure~\ref{fig:eff_vh4b} as a function of $|d_0|$ and $R_{\mathrm{prod}}$.
The performance of the primary and LRT algorithms are compared together with
the combination of the two.
After $|d_0| > 5$~mm, the primary track reconstruction efficiency becomes significantly reduced, with LRT
recovering the loss in efficiency with appreciable efficiencies out to $|d_0| = 300$~mm.
This translates in a total efficiency of about 40\% up to production radii of $R_{\mathrm{prod}} = 300$~mm.
A comparison of the combined track reconstruction efficiencies obtained for the three signal models considered is shown in Figure~\ref{fig:eff_tot_llp}.
As expected, the combined efficiency degrades for dense environments in a given $|d_0|$ bin.  In the region $10 <
|d_0| < 20$~mm, the total track reconstruction efficiency is approximately 85\%, 60\% and 40\% for heavy neutral lepton,
the Higgs sector and long-lived neutralino benchmark models, respectively.

\subsection{Robustness against pileup}
 
\begin{figure}[!htb]
\centering
\subfloat[]{\includegraphics[width=.45\textwidth]{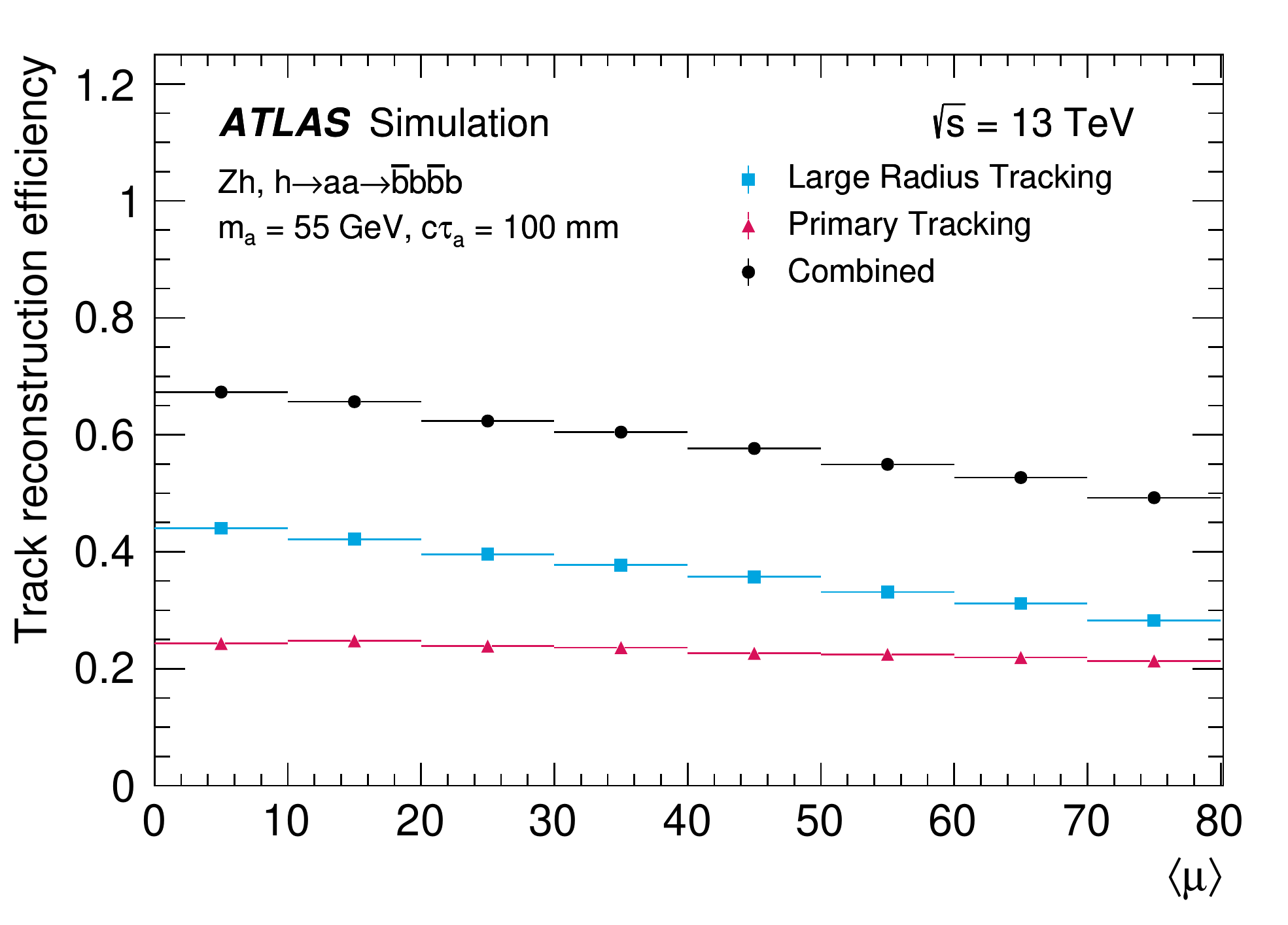}}
\qquad
\subfloat[]{\includegraphics[width=.45\textwidth]{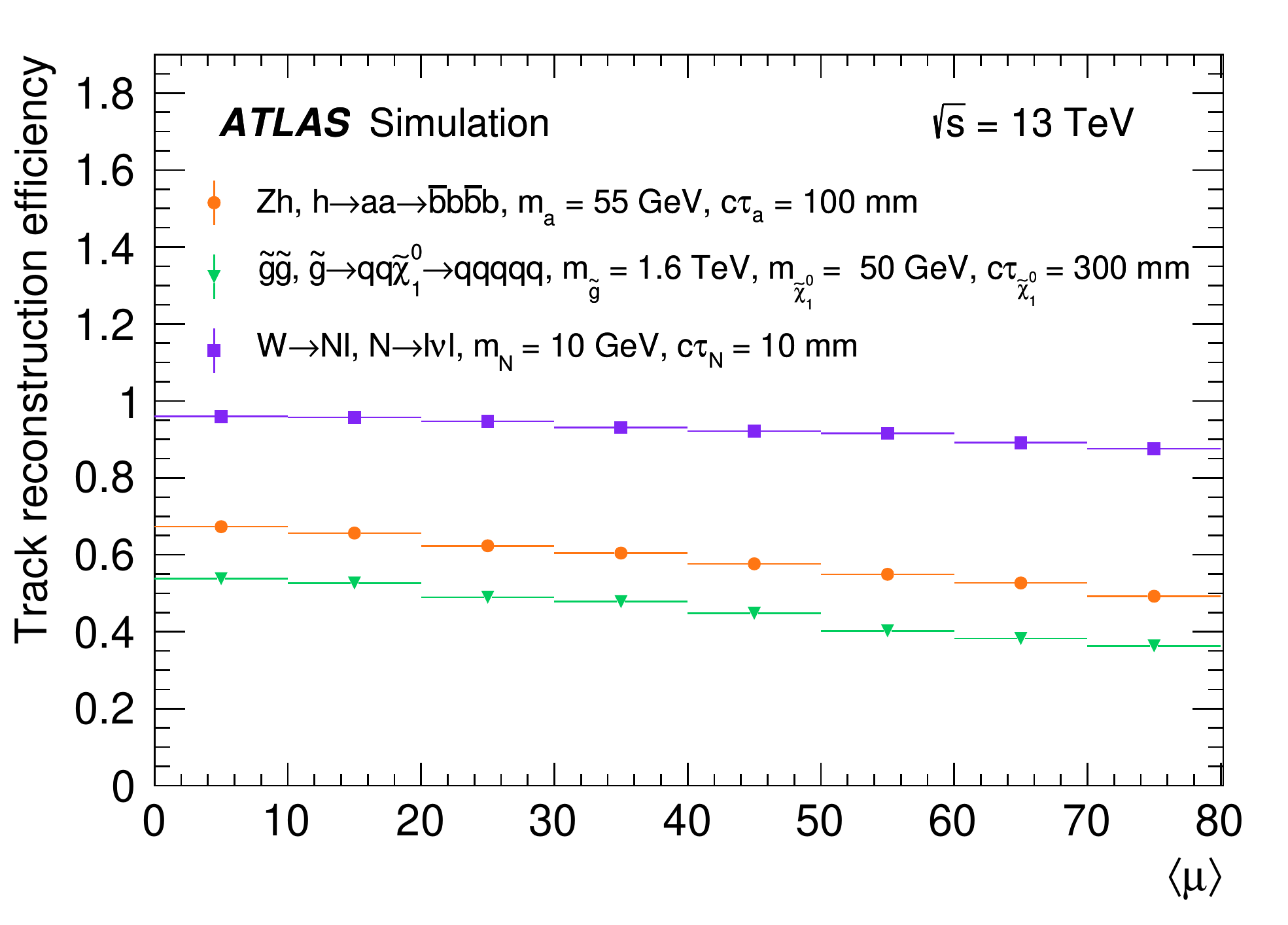}}
\caption{Track reconstruction efficiencies as a function of $\langle \mu \rangle$
for displaced charged particles produced by the decay of long-lived
signal particles.  (a) The primary, LRT and combined tracking
efficiencies are compared for the Higgs portal model. (b) The combined
efficiencies for the long-lived neutralino, Higgs portal and heavy neutral
lepton signal models.
Truth particles are required to fulfill the fiducial selection criteria stated in the text.
}
\label{fig:eff_mu}
\end{figure}

As the average number of interactions per bunch crossing increases, so does the
number of charged particles produced in an event.  This increase in
combinatorics can degrade the track reconstruction efficiency. 
The primary, LRT and combined track reconstruction efficiencies for the Higgs portal model are shown as a function of $\langle \mu \rangle$
in Figure~\ref{fig:eff_mu}. The comparison of the combined efficiencies of the three benchmark
signals is also provided showing more stable performance as a function of the
pileup for tracks originated from leptonic decays. This effect is primarily due to the lack of hadronic interactions for leptonic signatures.
 
\subsection{Impact on searches for displaced vertices}
 
To evaluate the impact of the improved LRT configuration on searches for
displaced vertices, it is useful to study how the updated track reconstruction
changes the LLP vertex reconstruction efficiency. Additionally, it is important to investigate the overall rate of
vertices reconstructed from fake tracks, which correspond to a dominant
background in such searches. The study is performed using the Higgs portal
benchmark model. 
To compare the performance between the two LRT configurations, the simulated events are reconstructed
using the legacy and updated ATLAS track reconstruction configurations.  Secondary
vertices are reconstructed using the inclusive vertexing algorithm described in
Section~\ref{sec:vertexing}. Identical vertex reconstruction configurations are
used for the two samples.
 
\begin{figure}[!htb]
\centering
\includegraphics[width=0.6\linewidth]{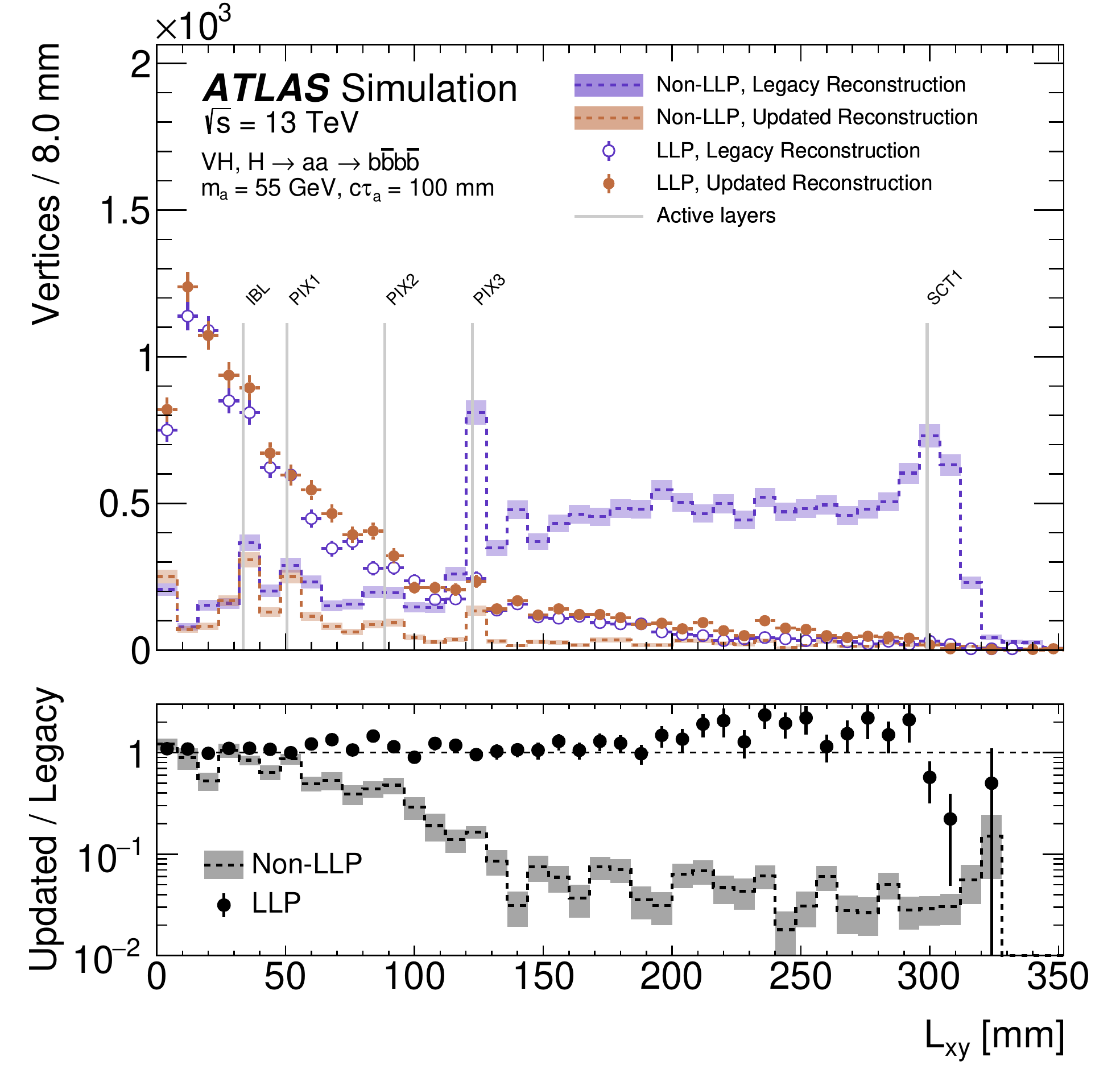}
\caption{A comparison of the radial distributions of reconstructed secondary
vertices in a simulated LLP sample using the legacy and updated
track reconstruction configurations. The circular markers represent
reconstructed vertices that are matched to truth-level LLP decay vertices
(LLP). The dashed lines represent reconstructed vertices that are not matched
to truth-level LLP decay vertices (non-LLP). The radial position of the pixel
and SCT detector layers are indicated by vertical lines. The lower panel shows
the ratio of the number of vertices reconstructed in the updated reconstruction to
those reconstructed in the legacy implementation for true LLP vertices (circular markers)
and non-LLP vertices (dashed lines). The uncertainties shown are statistical only.}
\label{fig:VSI}
\end{figure}

To assess the signal vertex reconstruction efficiency and evaluate the impact of
fake tracks, a truth-matching procedure is performed.  Reconstructed vertices
are considered matched to an LLP decay if the $p_{\mathrm{T}}$-weighted fraction
of tracks in the vertex that are matched to truth-level LLP descendants (as
described in Section~\ref{sec:fakes}) is greater than 0.5. Figure~\ref{fig:VSI}
shows the transverse displacement ($L_{xy}$) of truth-matched and non-truth-matched vertices
for the legacy and improved reconstruction. The improved reconstruction has a modest increase
in the number of LLP vertices and considerably fewer vertices not matched to LLP decays, non-LLP, due to the
significant increase in purity of the LRT algorithm.
 
\section{Data and simulation comparison}
\label{sec:data_mc_compare}

To further validate the improved LRT configuration, comparisons are performed
between Run 2 data and simulated inelastic $pp$ collisions.  As mentioned in
Section~\ref{sec:samples}, the simulation is generated with a flat $\langle \mu
\rangle$ profile and is thus reweighted to agree with the $\langle \mu \rangle$
profile of the data. The transverse momentum distributions of tracks are found
to be consistent between data and simulation therefore no additional
reweighting is performed. Track-level comparisons are performed
(Section~\ref{sec:trk}), as well as comparisons of the rate of fake
reconstructed tracks (Section~\ref{sec:fake}) and of the properties of secondary vertices consistent
with the decays of $K_S^0$ mesons (Section~\ref{sec:KS}).
 
\subsection{Track properties}
\label{sec:trk}
 
\begin{figure}[!h] \centering
\subfloat[]{\includegraphics[width=0.45\linewidth]{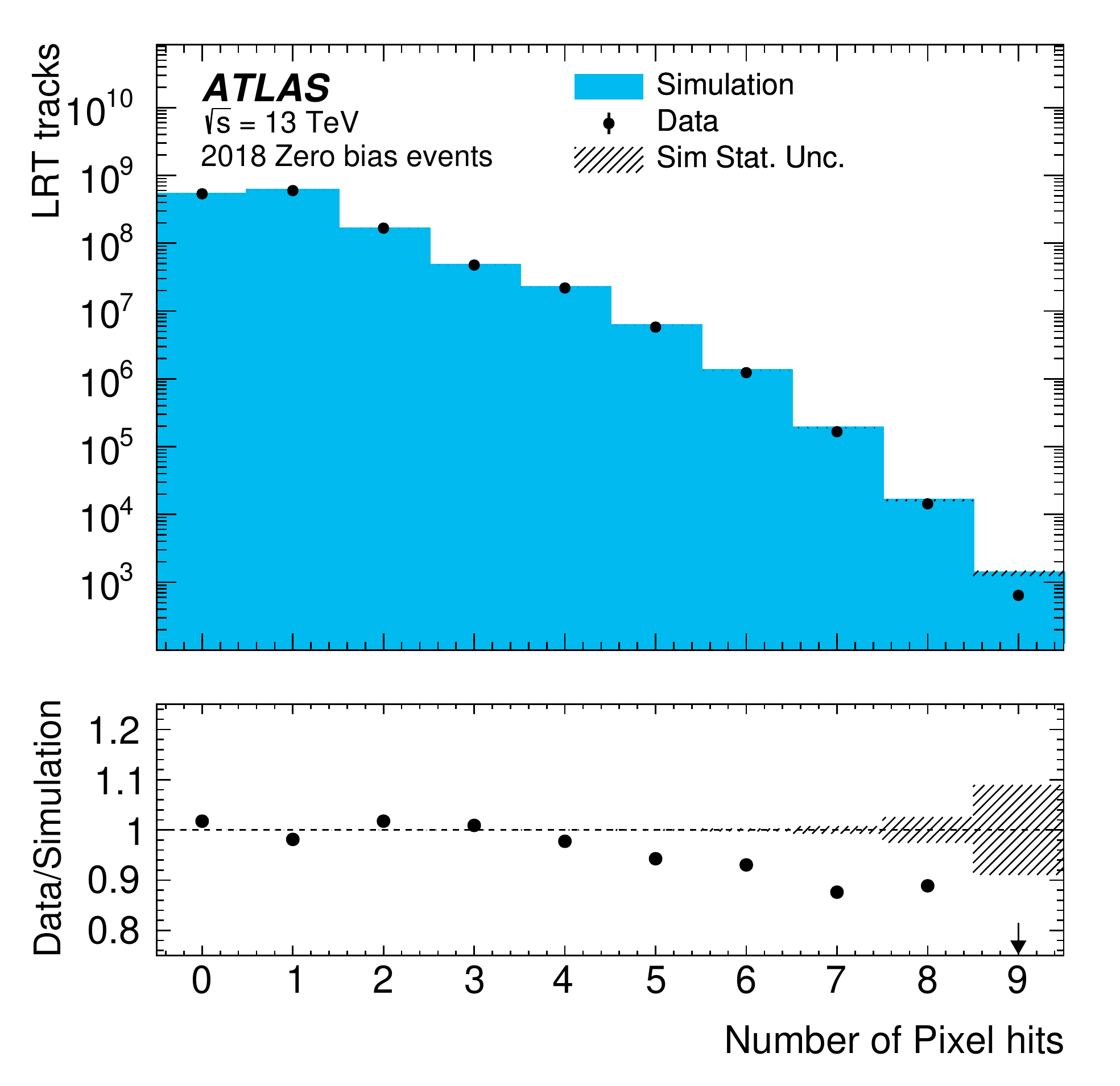}} \qquad
\subfloat[]{\includegraphics[width=0.45\linewidth]{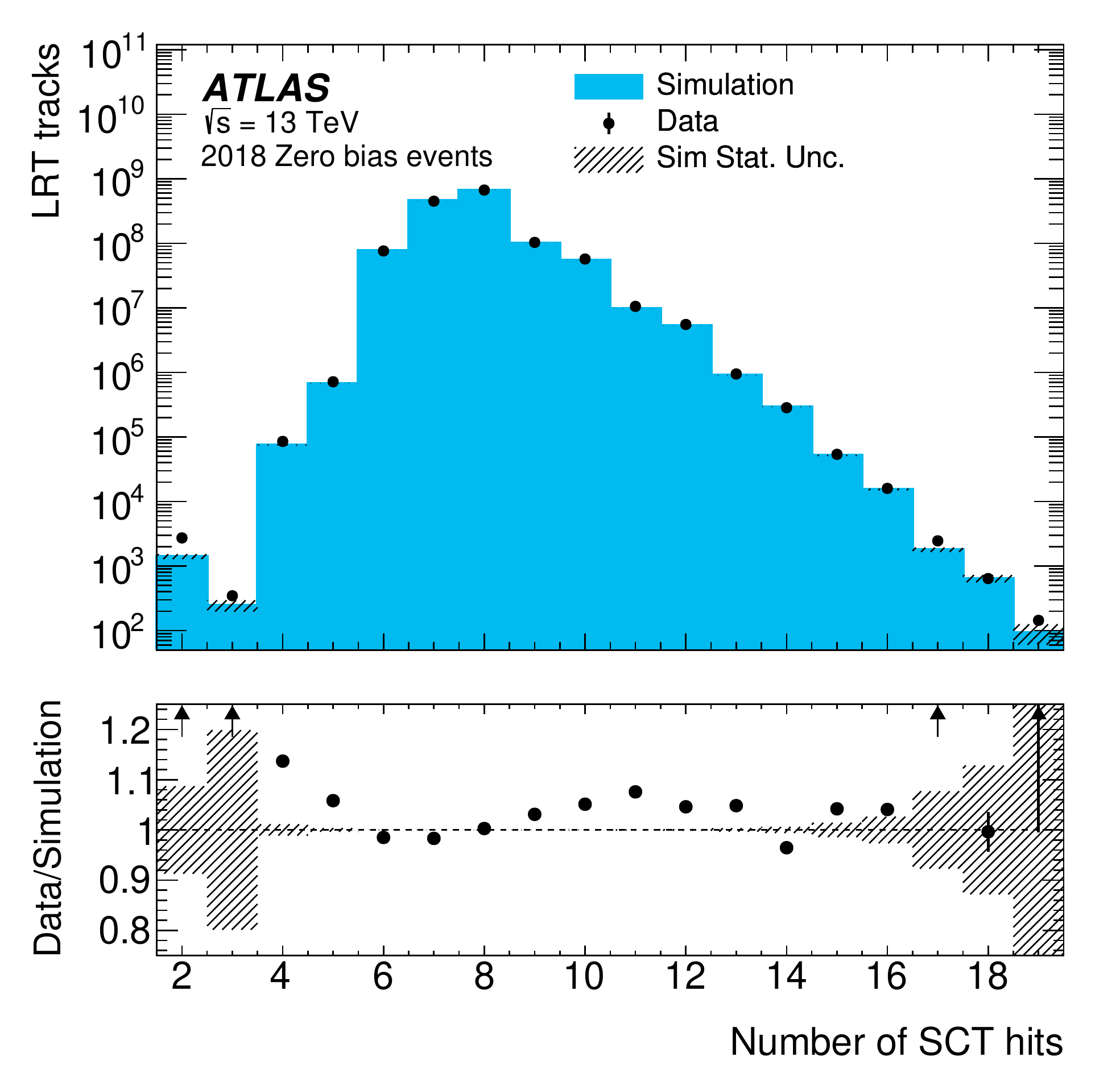}} \\
\subfloat[]{\includegraphics[width=0.45\linewidth]{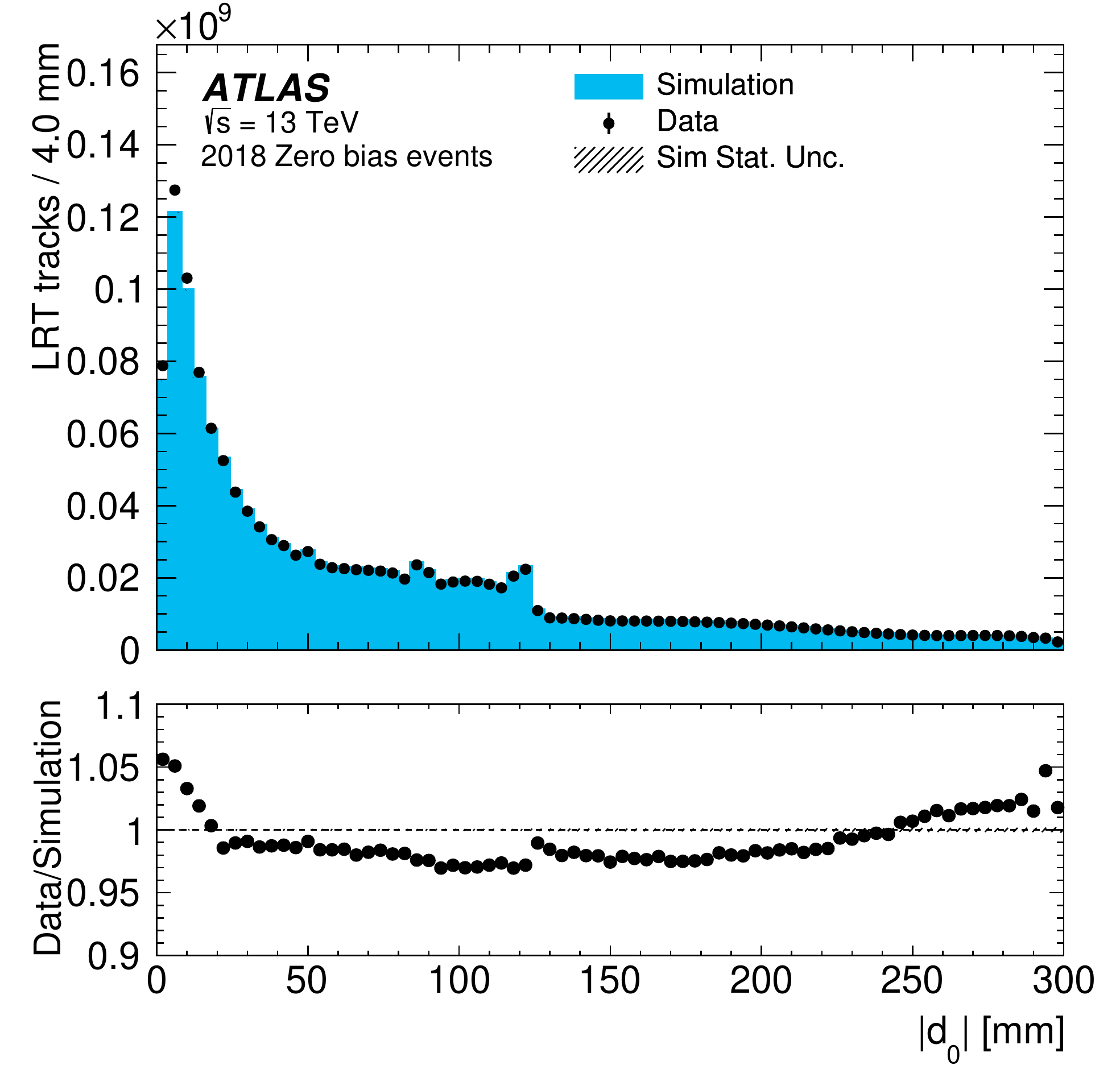}} \qquad
\subfloat[]{\includegraphics[width=0.45\linewidth]{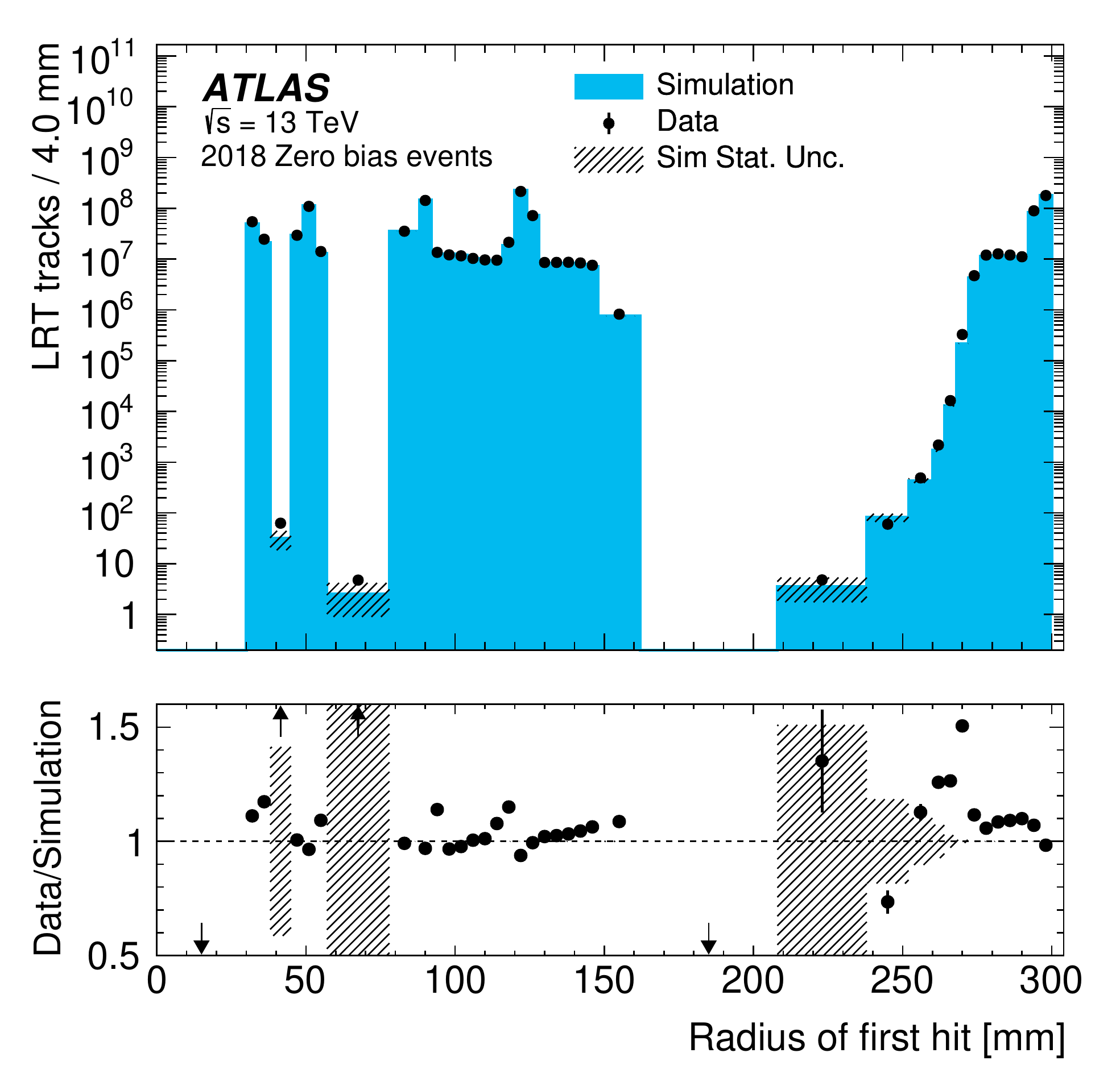}}
\caption{The distribution of the number of hits in the (a) pixel and (b) SCT
detector, (c) $|d_0|$, and (d) the radius of the first hit on each track, in 2018
zero bias data and simulated inelastic $pp$ collisions. The simulation is
normalised to the number of tracks in data. The statistical uncertainty on the
data and simulation are shown as vertical error bars and shaded bands, respectively.
Points in the ratio that do not lie within the visible range are shown as
arrows. }
\label{fig:hits}
\end{figure}

The distributions of the number of hits in the pixel and SCT ID subsystems per
LRT track are shown in Figure~\ref{fig:hits}, along with the distributions of
$|d_0|$ and the radius of the first hit associated to the track. The largest
deviations from unity in the ratio of data to simulation are at the $10\%$
level. This is consistent with the performance of the Run 2 LRT
configuration~\cite{ATL-PHYS-PUB-2017-014} and indicates good agreement between
data and simulation. The peaks in the distribution of $|d_0|$ at 50 mm, 88 mm, and 122 mm
coincide with the radial positions of the three pixel barrel layers, and
correspond to low-$p_{\mathrm{T}}$ secondary tracks with trajectories tangential
to the detector modules.
 
\subsection{Fake rate}
\label{sec:fake}
 
\begin{figure}[!h]
\centering
\subfloat[]{\includegraphics[width=0.45\linewidth]{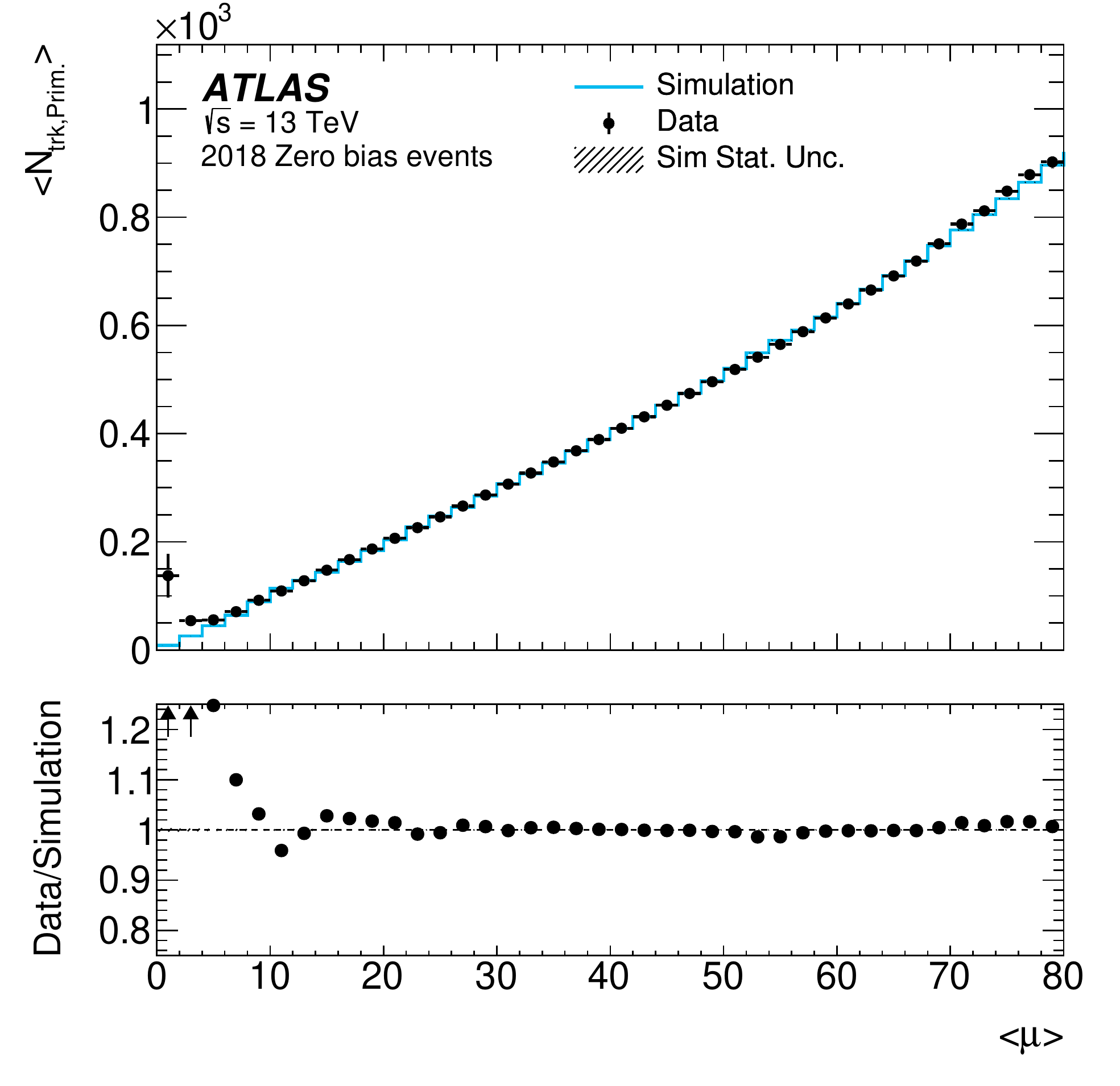}}
\qquad
\subfloat[]{\includegraphics[width=0.45\linewidth]{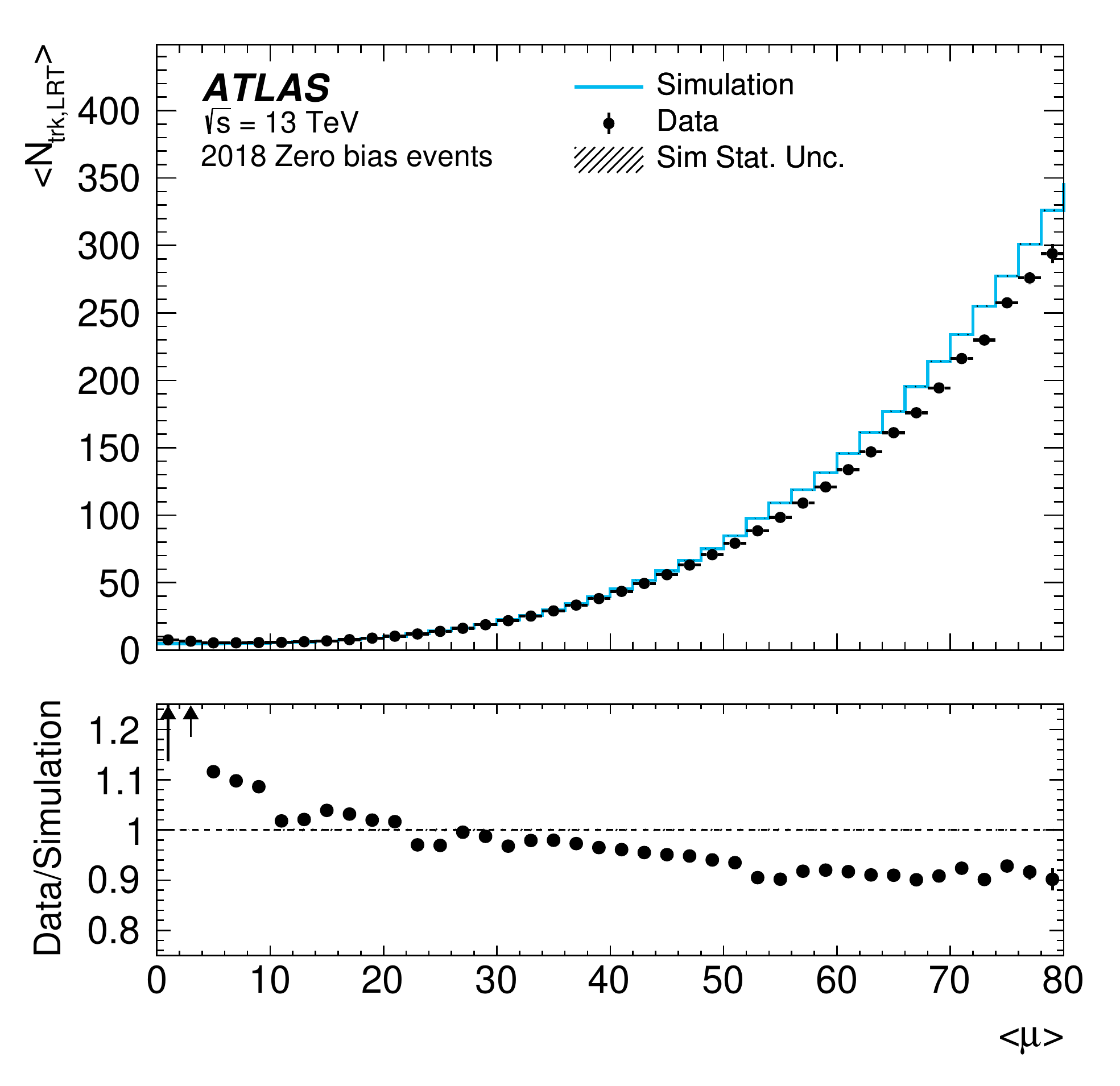}}
\caption{The mean number of tracks per event as a function of $\langle \mu \rangle$
for (a) primary tracks and (b) LRT tracks.  The simulation is scaled by extracting a scale factor from the ratio of
$\langle N_{\mathrm{track,Prim.}} \rangle$ of data to simulation.}
\label{fig:ntrk_vs_mu}
\end{figure}

The number of tracks corresponding to real charged particle
trajectories is expected to scale linearly with $\langle \mu \rangle$, since it
is related to the number of charged particles produced in the collisions.
However, during reconstruction, it is possible for the algorithm to reconstruct
fake tracks not corresponding to real charged particle trajectories. The number
of these random combinations is expected to scale with a higher power as a
function of $\langle \mu \rangle$, since the increased combinatorics allow for
more track candidates to be formed.  The average number of primary tracks per
event ($\langle N_{\mathrm{track,Prim.}} \rangle$) and the average number of
LRT tracks per event ($\langle N_{\mathrm{track,LRT}} \rangle$) as a
function of $\langle \mu \rangle$ is shown in Figure~\ref{fig:ntrk_vs_mu} for
data and simulation. To account for differences between the data and simulation
in terms of the overall number of charged particles per event, an overall
normalisation factor is used to adjust the number of tracks in simulation to
data. This scale factor is determined to be 0.94, and is applied to simulation
in both distributions.
 
Comparing the distributions of $\langle N_{\mathrm{track,Prim.}} \rangle$ to
$\langle N_{\mathrm{track,LRT}} \rangle$, it is clear that the fraction of fake
tracks (deviation from linearity with $\langle \mu \rangle$) for the LRT reconstruction is significantly larger than for primary
tracking. However, the overall number of tracks reconstructed by the LRT
reconstruction is small compared to the number of primary tracks, indicating
that the impact of the LRT fake rate is minimal to the overall size of the final
track collection.
Additionally, the difference between the measured value of $\langle N_{\mathrm{track,LRT}} \rangle$ in simulation
and data grows with increasing $\langle \mu \rangle$, indicating that the rate of fake tracks is larger in
simulation than in data. This difference is driven by the modeling of the detector material in the simulation
and will be discussed further in Section~\ref{sec:syst}.
 
\subsection{Study of $K_S^0$ vertices}
\label{sec:KS}
 
\begin{figure}[!htb]
\centering
\subfloat[]{\label{fig:Ks_mass_STD}\includegraphics[width=0.45\linewidth]{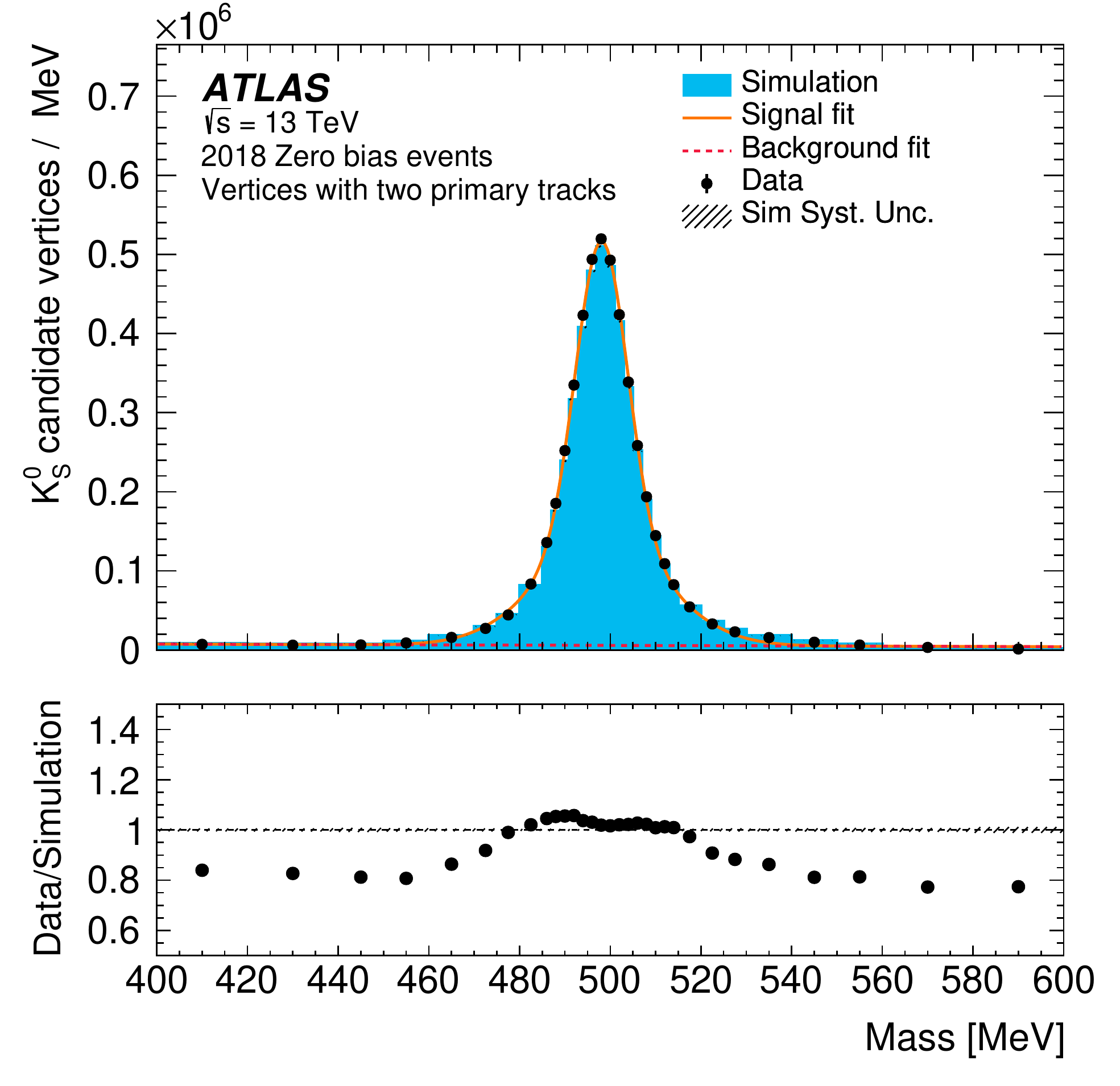}}
\subfloat[]{\label{fig:Ks_mass_LRT}\includegraphics[width=0.45\linewidth]{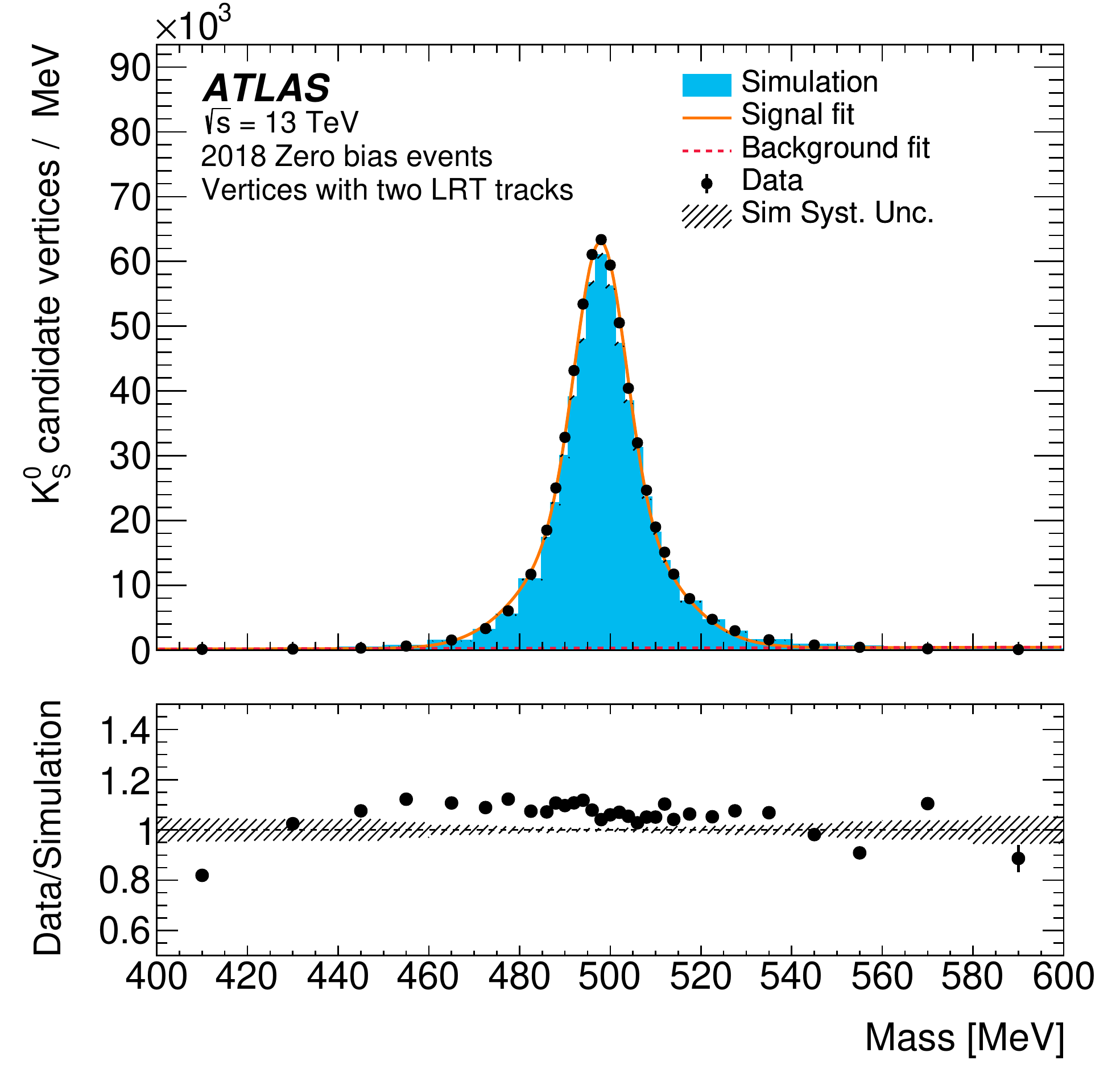}} \\
\caption{The invariant mass of $K_S^0$ candidates reconstructed from (a) two
primary tracks and (b) two LRT
tracks, in 2018 zero bias data and
simulated inelastic $pp$ collisions. The signal peak is fitted to a
double-gaussian, and a third-order polynomial is used to model the
combinatorial background. The yields in simulation are normalised such that
the number of $K_S^0$ candidates with two primary tracks agrees with the
data.}
\label{fig:Ks_mass}
\end{figure}

To compare the track reconstruction efficiency between data and simulation, a
``standard candle'' is needed to identify LRT tracks originating from a true LLP
decay.  The $K_S^0$ meson is an ideal candidate due to its proper decay length
($c\tau = 27$~mm) and a clean signature of a two-track vertex. To identify
$K_S^0$ vertices, secondary vertices are reconstructed from the combined
collection of primary and LRT tracks using the algorithm designed for
reconstructing $V0$ decays described in Section~\ref{sec:vertexing}. $K_S^0$
candidates are identified by requiring that the combined invariant mass of the
two selected tracks is in the range [472, 522] MeV. No vetoes or background
subtraction is applied to the vertices beyond those applied during
reconstruction, as described in Section~\ref{sec:vertexing}. The $\pt$
distributions of the $K_S^0$ candidates were found to be well modeled in the
simulation and thus no $\pt$ re-weighting is applied. The distributions of the
invariant mass of the $K_S^0$ candidates are shown for data and simulation in
Figure~\ref{fig:Ks_mass} for vertices with two primary tracks and two large
radius tracks. The yields in simulation are normalised such that the number of
$K_S^0$ candidates with two primary tracks agrees with the data.  To assess the
purity of the $K_S^0$ selection, the signal peak is fitted to a double-gaussian,
and a third-order polynomial is used to extract the combinatorial background.
Both distributions show very small combinatorial backgrounds with purities of
roughly 99\%.
 
\begin{figure}[!ht]
\centering
\subfloat[]{\includegraphics[width=0.45\linewidth]{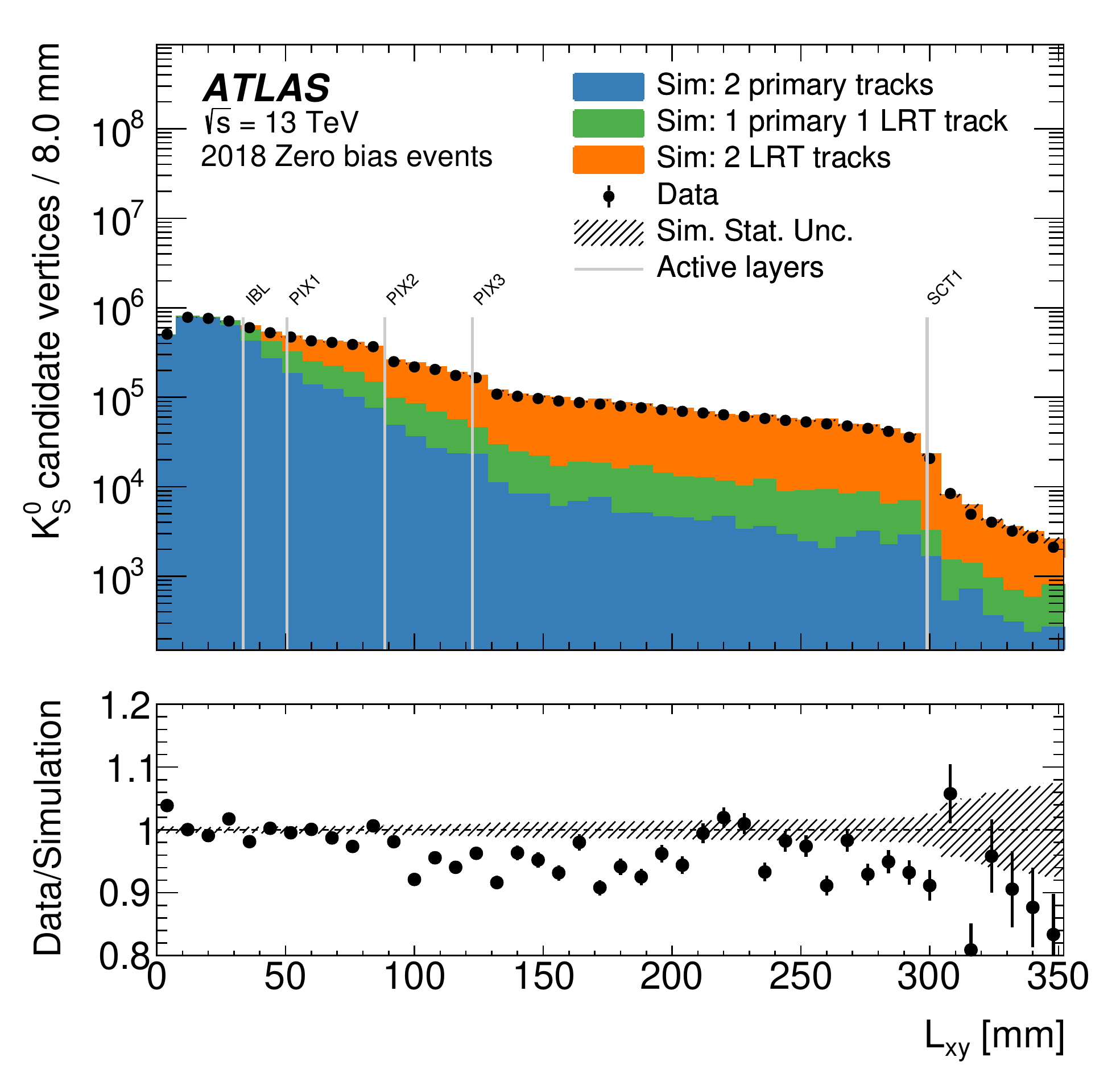}\label{fig:KS_r}}
\qquad
\subfloat[]{\includegraphics[width=0.45\linewidth]{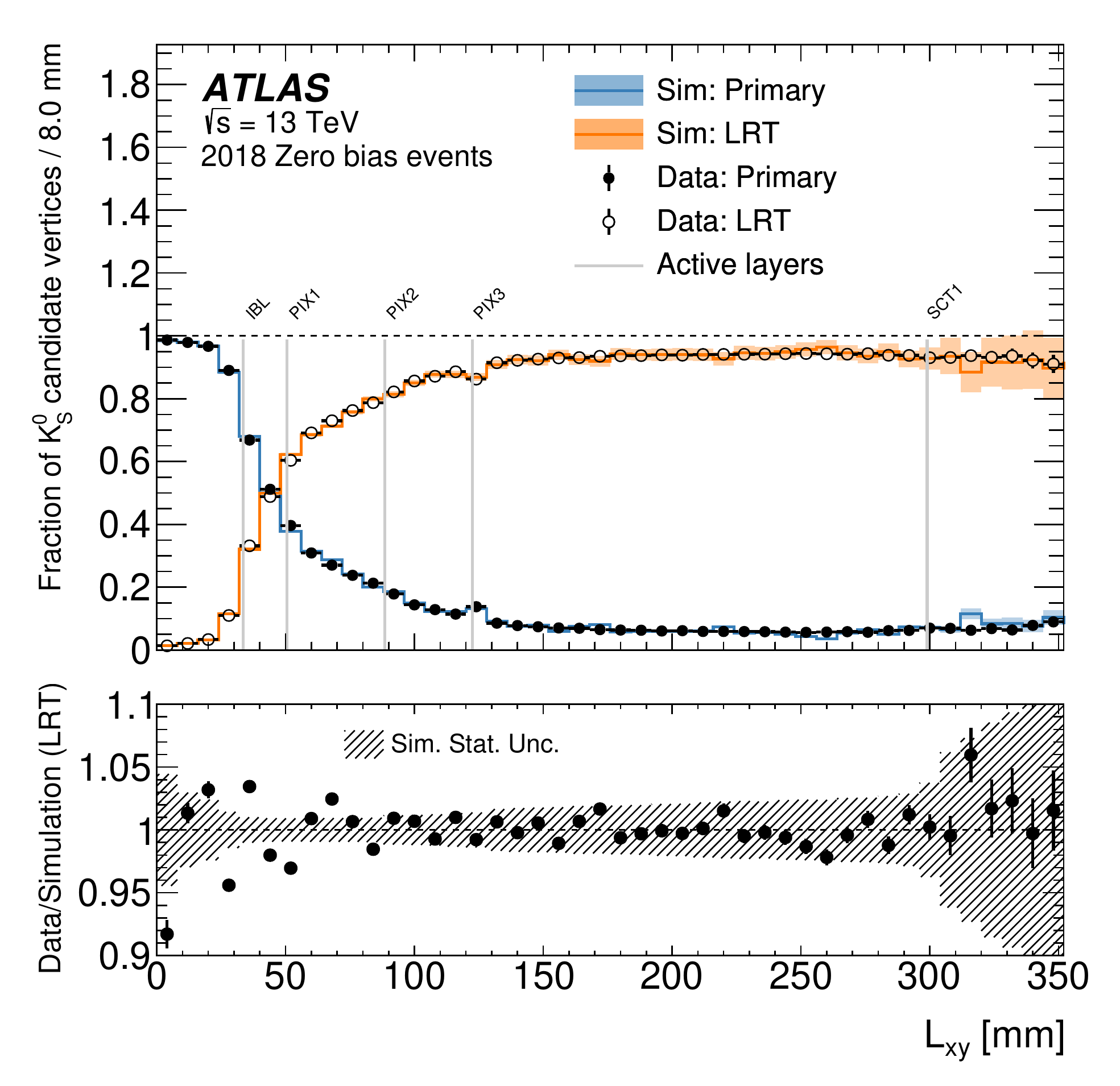}\label{fig:KS_r_ratio}}
\caption{\protect\subref{fig:KS_r} The radial distribution of reconstructed
$K_S^0$ candidate vertices in 2018 zero bias data and
simulated inelastic $pp$ collisions. The yields
in simulation are normalised such that the number of $K_S^0$ candidates with
two primary tracks agrees with the data. The lower panel displays the ratio
between the total number of reconstructed $K_S^0$ candidate vertices in the
data and the simulation. The statistical uncertainty on the simulation is
shown as a hatched band.
\protect\subref{fig:KS_r_ratio} The fraction of $K_S^0$ candidate vertices
reconstructed with two primary tracks (labelled ``Primary'' in the legend) and
at least one large radius track (labelled ``LRT'' in the legend) in 2018 zero
bias data and simulated inelastic $pp$ collisions.  The four pixel layers and
first SCT layer are shown in the figure with gray vertical lines.  The lower
panel displays the ratio between the fraction of reconstructed $K_S^0$
candidate vertices with at least one LRT track in the data and the simulation.
}
\label{fig:KS}
\end{figure}

The radial distribution ($L_{xy}$) of the reconstructed $K_S^0$ candidates are
then compared between data and simulation, as shown in Figure~\ref{fig:KS_r}.
Vertices in simulation are separated in three categories, based on the number of
LRT tracks included in the vertex fit.
After the first SCT layer, LRT begins to have a diminishing contribution to the
total yield. This is due to the fact that particles produced in this region will
traverse an insufficient number of silicon layers to satisfy the hit requirements of
inside-out track reconstruction and the primary outside-in tracking becomes dominant.
Figure~\ref{fig:KS_r_ratio} shows the fraction of $K_S^0$ candidate vertices
reconstructed with either two primary tracks or with at least one LRT
track in data and simulation.  The distributions are consistent between the two
samples, further indicating good modeling of the LRT reconstruction in
simulation. After $L_{xy} > 70$~mm, LRT tracks account for more than
75\% of the total number of $K_S^0$ candidate vertices, highlighting the
importance of LRT for the reconstruction of displaced vertices.
 
\subsubsection{Systematic uncertainty on the LRT reconstruction efficiency}
\label{sec:syst}
 
In Figure~\ref{fig:Ks_mass_LRT}, a small difference can be observed between the
number of $K_S^0$ candidates reconstructed with two LRT tracks between
data and simulation, despite normalising the simulation such that a consistent
number of candidates are reconstructed with two primary tracks. This indicates
that there is a systematic difference between the LRT reconstruction efficiency
measured in data and simulation. To account for this in physics analyses, a
per-track uncertainty on the LRT reconstruction efficiency is derived by
comparing the yields of LRT tracks associated to $K_S^0$ candidates as a
function of $|\eta|$ and the radius of the first measurement on the track
($R_{1^{\mathrm{st}} \, \mathrm{hit}}$).  A summary of the measured relative
uncertainties is provided in Table~\ref{tab:LRT_uncertainties}. To propagate
these uncertainties to downstream object reconstruction algorithms such as
secondary vertex reconstruction, a procedure is used in which tracks are
randomly removed from the track collection considered by the vertex
reconstruction with a probability corresponding to the per-track uncertainty of
each track. This results in a modified vertex collection which can be used to
assess the impact of the per-track uncertainties on the final vertex
reconstruction performance.
 
\begin{table}[!h]
\begin{center}
\caption{The relative per-track uncertainties on the LRT reconstruction
efficiency in various bins of $|\eta|$ and radius of the first measurement
on the track $R_{1^{\mathrm{st}} \, \mathrm{hit}}$.}

\makebox[0pt]{
\begin{tabular}{c c c c}
\toprule
$R_{1^{\mathrm{st}} \, \mathrm{hit}}$ [mm]               & $|\eta| < 0.4$ & $0.4 < |\eta| < 1.2$ & $1.2 < |\eta| < 3.0$ \\ \midrule
$[33.0, 38.0]$               &  16.8\%        &  11.3\%              &  7.2\%               \\
$[46.0, 56.0]$   					   &  10.0\%        &  4.3\%               &  5.6\%               \\
$[84.0, 94.0]$   					   &  0.3\%         &  1.6\%               &  2.8\%               \\
$[118.0, 128.0]$ 					   &  0.8\%         &  3.9\%               &  6.6\%               \\
$[288.0, 320.0]$ 					   &  9.6\%         &  0.4\%               &  1.1\%               \\ \bottomrule
\end{tabular}
}


\label{tab:LRT_uncertainties}
\end{center}
\end{table}

\begin{figure}[h!]
\centering
\subfloat[]{\includegraphics[width=0.45\linewidth]{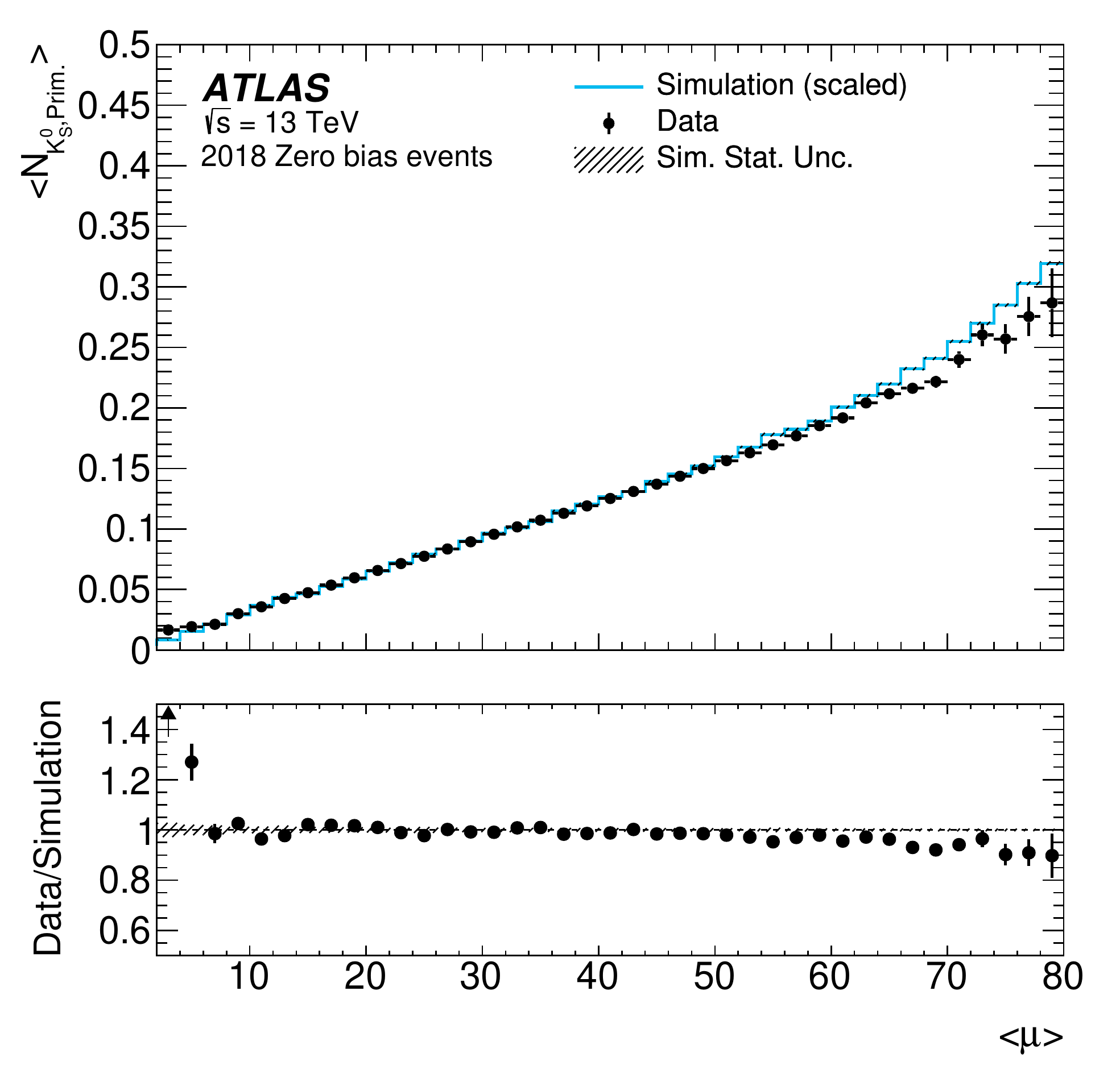}\label{fig:KS_mu_STD}}
\qquad
\subfloat[]{\includegraphics[width=0.45\linewidth]{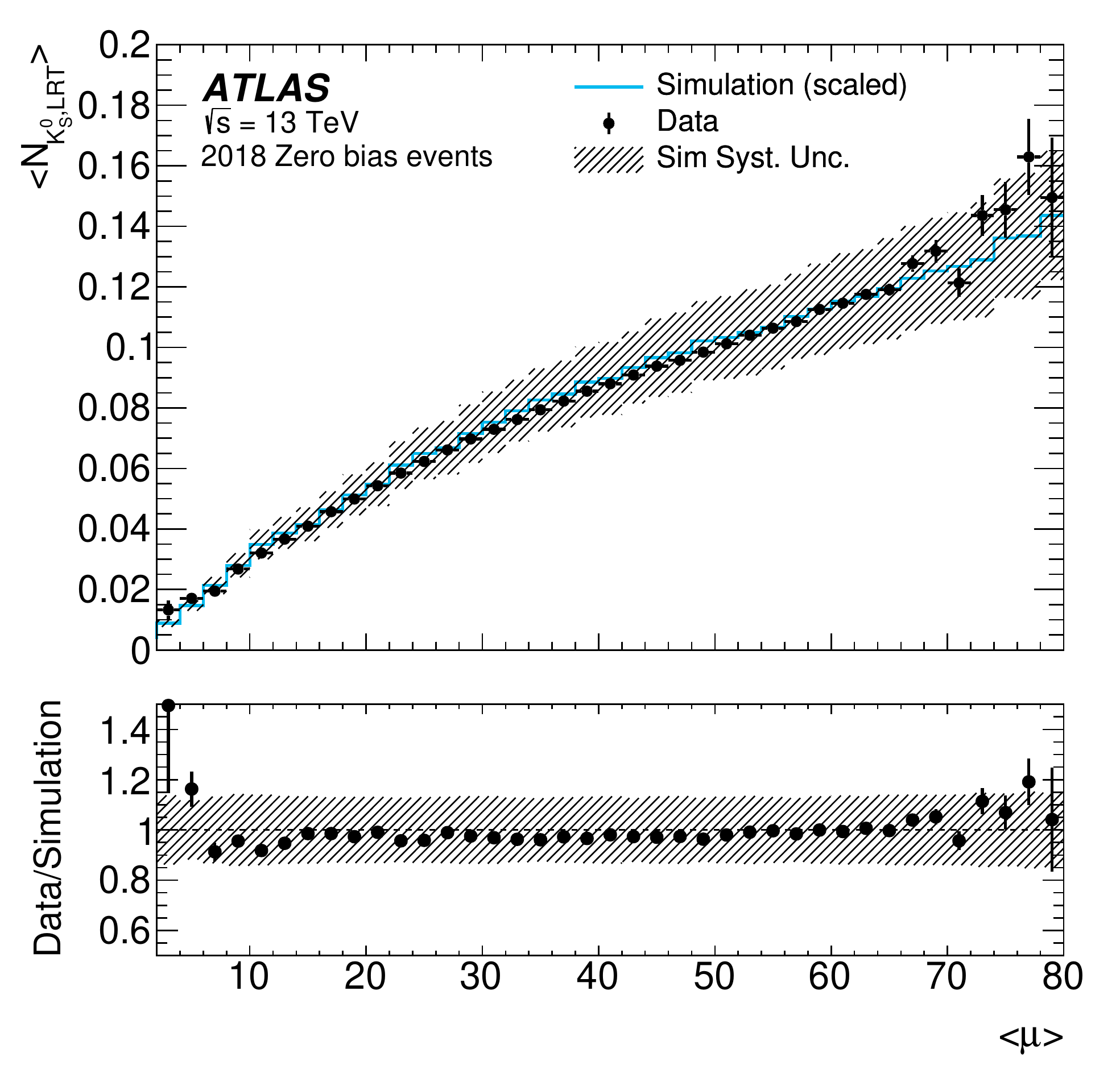}\label{fig:KS_mu_LRT}}
\caption{The average number of $K_S^0$ candidate vertices per event in 2018
zero bias data and simulated inelastic $pp$ collisions shown as a function of
the average number of interactions per bunch crossing $\langle \mu \rangle$
for (a) vertices with two primary tracks and (b) vertices with two LRT tracks.
The simulation is scaled such that the total number of reconstructed $K_S^0$
candidates with two primary tracks agrees with the data. The uncertainty on
the simulation in (b) is derived by propagating the per-track uncertainties
through to the vertex reconstruction using the method described in the text.}
\label{fig:KS_mu}
\end{figure}
 
The average number of $K_S^0$ candidate vertices per event is shown in
Figure~\ref{fig:KS_mu} as a function of $\langle \mu \rangle$ for vertices with
two primary tracks and two LRT tracks. In both cases, the simulation is scaled such that the total number of
reconstructed $K_S^0$ candidates with two primary tracks agrees with the data.
Similar to the distributions of $\langle
N_{\mathrm{trk}} \rangle$, this is expected to scale linearly if the track
reconstruction efficiency and proportion of background vertices stays constant
as a function of $\langle \mu \rangle$. However, as $\langle \mu \rangle$
increases, so does the overall number of tracks in the event. This increases the
probability for random track crossings that can mimic a $K_S^0$ candidate,
contributing to the deviation from linearity seen in Figure~\ref{fig:KS_mu_STD}.
For vertices with two LRT tracks however, Figure~\ref{fig:KS_mu_LRT} shows that
the number of $K_S^0$ candidate vertices starts to deviate downward from
linearity. This indicates that even though the number of LRT fake tracks is
high, the rate of fake vertices is relatively low.  The main factor affecting
the number of $K_S^0$ candidates at high $\langle \mu \rangle$ values is the
reduction in overall LRT reconstruction efficiency, as depicted in
Figure~\ref{fig:eff_mu}.  The difference between the number of $K_S^0$ candidate
vertices per event in data and simulation shown in Figure~\ref{fig:KS_mu_LRT} is
within the systematic uncertainties on the final vertex yields derived by
propagating the per-track LRT uncertainties to the vertex reconstruction, as
described above.
 
\begin{figure}
\centering
\includegraphics[width=0.6\linewidth]{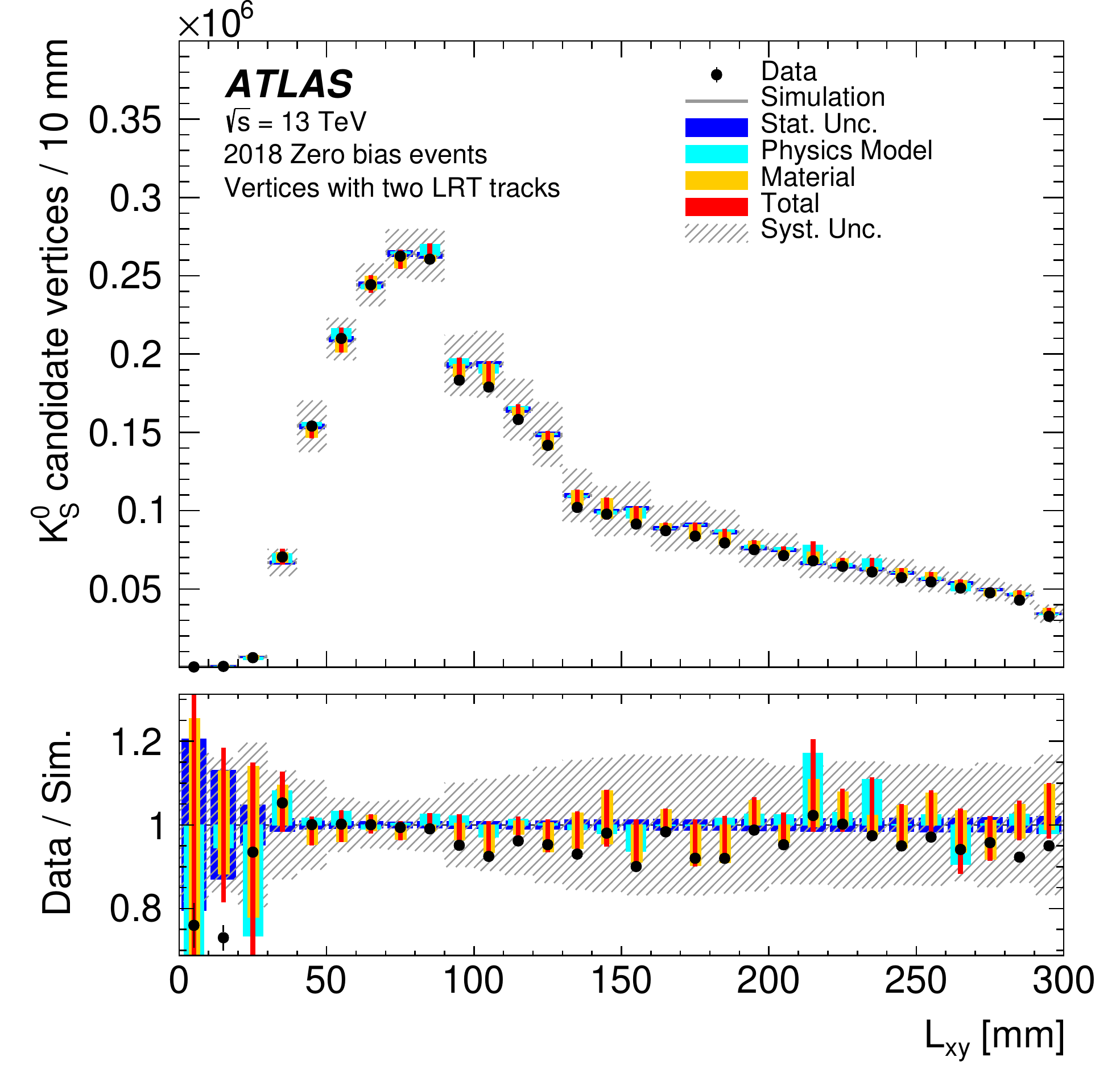}
\caption{The radial distribution of reconstructed $K_S^0$ candidate vertices
with two LRT tracks in 2018 zero bias data and simulated inelastic $pp$
collisions. The yields in simulation are normalised such that the number of
$K_S^0$ candidates with two primary tracks agrees with the data. Samples with
varied detector geometries and simulation parameters are used to derive
uncertainties on the nominal simulation and are shown as filled bands. The
uncertainties derived from each geometry variation are summed in quadrature to
obtain a single uncertainty which is shown as an orange band (labeled
"Material" in the legend). The uncertainty derived from comparing the nominal
simulation to the sample with the modified physics list is shown as a light blue
band (labeled "Physics Model" in the legend). The statistical uncertainty on
the nominal sample is shown as a dark blue band (labeled "Stat.  Unc." in the
legend). The three sources of uncertainty (material, physics model, and
statistical uncertainty) are then summed in quadrature to obtain a total
uncertainty on each bin which is drawn as a red band (labeled "Total"
in the legend). The systematic uncertainty derived following the method
described in the text is shown as a hatched band (labeled "Syst. Unc." in the
legend). The lower panel displays the ratio between the data and the nominal
simulation and the relative values of each uncertainty.}
\label{fig:Ks_R_LRT_material}
\end{figure}

The discrepancy in yields can be attributed to the modeling of the Inner
Detector material~\cite{PERF-2015-07}, and is explored further through an
analysis of the alternate simulated samples described in
Section~\ref{sec:samples}.  Three samples with modified detector geometries are
considered: an overall 5\% increase of the passive material in the ID, a 10\%
increase of the material in the IBL, and a 25\% increase of the material
describing the pixel services.  A fourth sample is considered which uses the
QGSP\_BIC physics list instead of the nominal FTFP\_BERT physics list in the
GEANT4 simulation. Each simulated sample is first normalized such that the
number of simulated events is consistent with the data.  An additional
normalisation factor is then applied to each sample, which is defined such that
after normalisation the number of $K_S^0$ candidates with two primary tracks in
the nominal sample agrees with the data. The differences in yields between the
nominal sample and the varied samples are then computed and used to derive an
uncertainty band corresponding to each variation. The results of this analysis
are shown in Figure~\ref{fig:Ks_R_LRT_material}. The data yields are observed to
fall within the range of values obtained from the set of varied simulated
samples, indicating that the difference in LRT tracking efficiency between the
data and the simulation is well characterised by these variations.
Additionally, the systematic uncertainty derived by propagating the per-track
LRT uncertainties to the vertex reconstruction is shown, and is found to cover
the observed differences between the data and the simulation as well as the
variations between the alternate simulated samples.
 
\FloatBarrier
 
\section{Conclusion}
\label{sec:conclusion}

 
The ATLAS LRT algorithm has undergone significant improvements in preparation
for Run 3 data taking. As a result of this reoptimisation, the processing time
has improved by more than a factor of 10 and the disk space usage per event for
LRT tracks has been reduced by more than a factor of 50 as compared to the
legacy implementation, while still maintaining comparable track reconstruction
efficiency.  These improvements have allowed for LRT to be integrated in the
standard ATLAS reconstruction chain. This will significantly simplify the
workflow for future LLP analyses as well as allow for analyses not
specifically targeting LLP signatures to benefit from increased track
reconstruction efficiency at large $|d_0|$. A study of secondary vertex
reconstruction for hadronic LLP signatures indicates that the improved LRT
configuration leads to a significant reduction in vertices reconstructed from
fake tracks and an overall improvement in the vertex reconstruction efficiency.
This will translate directly to an improved signal to background ratio for
displaced vertex searches in Run 3. To benefit from the improved performance,
Run 2 data events have been reconstructed using these updated configurations.
Additionally, comparisons between data and simulation show good modelling of both low-level track quantities as well as
high-level properties of reconstructed $K_S^0$ vertices using LRT
tracks. These developments are expected to increase the sensitivity of LLP
searches in Run 3 and allow for a considerable expansion of the ATLAS LLP search
program.


\FloatBarrier
\section*{Acknowledgments}


We thank CERN for the very successful operation of the LHC, as well as the
support staff from our institutions without whom ATLAS could not be
operated efficiently.
 
We acknowledge the support of
ANPCyT, Argentina;
YerPhI, Armenia;
ARC, Australia;
BMWFW and FWF, Austria;
ANAS, Azerbaijan;
CNPq and FAPESP, Brazil;
NSERC, NRC and CFI, Canada;
CERN;
ANID, Chile;
CAS, MOST and NSFC, China;
Minciencias, Colombia;
MEYS CR, Czech Republic;
DNRF and DNSRC, Denmark;
IN2P3-CNRS and CEA-DRF/IRFU, France;
SRNSFG, Georgia;
BMBF, HGF and MPG, Germany;
GSRI, Greece;
RGC and Hong Kong SAR, China;
ISF and Benoziyo Center, Israel;
INFN, Italy;
MEXT and JSPS, Japan;
CNRST, Morocco;
NWO, Netherlands;
RCN, Norway;
MEiN, Poland;
FCT, Portugal;
MNE/IFA, Romania;
MESTD, Serbia;
MSSR, Slovakia;
ARRS and MIZ\v{S}, Slovenia;
DSI/NRF, South Africa;
MICINN, Spain;
SRC and Wallenberg Foundation, Sweden;
SERI, SNSF and Cantons of Bern and Geneva, Switzerland;
MOST, Taiwan;
TENMAK, T\"urkiye;
STFC, United Kingdom;
DOE and NSF, United States of America.
In addition, individual groups and members have received support from
BCKDF, CANARIE, Compute Canada and CRC, Canada;
PRIMUS 21/SCI/017 and UNCE SCI/013, Czech Republic;
COST, ERC, ERDF, Horizon 2020 and Marie Sk{\l}odowska-Curie Actions, European Union;
Investissements d'Avenir Labex, Investissements d'Avenir Idex and ANR, France;
DFG and AvH Foundation, Germany;
Herakleitos, Thales and Aristeia programmes co-financed by EU-ESF and the Greek NSRF, Greece;
BSF-NSF and MINERVA, Israel;
Norwegian Financial Mechanism 2014-2021, Norway;
NCN and NAWA, Poland;
La Caixa Banking Foundation, CERCA Programme Generalitat de Catalunya and PROMETEO and GenT Programmes Generalitat Valenciana, Spain;
G\"{o}ran Gustafssons Stiftelse, Sweden;
The Royal Society and Leverhulme Trust, United Kingdom.
 
The crucial computing support from all WLCG partners is acknowledged gratefully, in particular from CERN, the ATLAS Tier-1 facilities at TRIUMF (Canada), NDGF (Denmark, Norway, Sweden), CC-IN2P3 (France), KIT/GridKA (Germany), INFN-CNAF (Italy), NL-T1 (Netherlands), PIC (Spain), ASGC (Taiwan), RAL (UK) and BNL (USA), the Tier-2 facilities worldwide and large non-WLCG resource providers. Major contributors of computing resources are listed in Ref.~\cite{ATL-SOFT-PUB-2021-003}.


\printbibliography
 
\clearpage
 
\input{atlas_authlist}

\end{document}

%% file: atlas_authlist.tex
 
\begin{flushleft}
\hypersetup{urlcolor=black}
{\Large The ATLAS Collaboration}

\bigskip

\AtlasOrcid[0000-0002-6665-4934]{G.~Aad}$^\textrm{\scriptsize 102}$,
\AtlasOrcid[0000-0002-5888-2734]{B.~Abbott}$^\textrm{\scriptsize 120}$,
\AtlasOrcid[0000-0002-1002-1652]{K.~Abeling}$^\textrm{\scriptsize 55}$,
\AtlasOrcid[0000-0001-5763-2760]{N.J.~Abicht}$^\textrm{\scriptsize 49}$,
\AtlasOrcid[0000-0002-8496-9294]{S.H.~Abidi}$^\textrm{\scriptsize 29}$,
\AtlasOrcid[0000-0002-9987-2292]{A.~Aboulhorma}$^\textrm{\scriptsize 35e}$,
\AtlasOrcid[0000-0001-5329-6640]{H.~Abramowicz}$^\textrm{\scriptsize 151}$,
\AtlasOrcid[0000-0002-1599-2896]{H.~Abreu}$^\textrm{\scriptsize 150}$,
\AtlasOrcid[0000-0003-0403-3697]{Y.~Abulaiti}$^\textrm{\scriptsize 117}$,
\AtlasOrcid[0000-0003-0762-7204]{A.C.~Abusleme~Hoffman}$^\textrm{\scriptsize 137a}$
% ...
\bigskip
\\

% ...
$^{29}$Physics Department, Brookhaven National Laboratory, Upton NY; United States of America.\\
% ...
$^{35}$$^{(a)}$Facult\'e des Sciences Ain Chock, R\'eseau Universitaire de Physique des Hautes Energies - Universit\'e Hassan II, Casablanca;$^{(b)}$Facult\'{e} des Sciences, Universit\'{e} Ibn-Tofail, K\'{e}nitra;$^{(c)}$Facult\'e des Sciences Semlalia, Universit\'e Cadi Ayyad, LPHEA-Marrakech;$^{(d)}$LPMR, Facult\'e des Sciences, Universit\'e Mohamed Premier, Oujda;$^{(e)}$Facult\'e des sciences, Universit\'e Mohammed V, Rabat;$^{(f)}$Institute of Applied Physics, Mohammed VI Polytechnic University, Ben Guerir; Morocco.\\
% ...
$^{49}$Fakult\"{a}t Physik , Technische Universit{\"a}t Dortmund, Dortmund; Germany.\\
% ...
$^{55}$II. Physikalisches Institut, Georg-August-Universit\"{a}t G\"ottingen, G\"ottingen; Germany.\\
% ...
$^{102}$CPPM, Aix-Marseille Universit\'e, CNRS/IN2P3, Marseille; France.\\
% ...
$^{117}$Department of Physics, New York University, New York NY; United States of America.\\
% ...
$^{120}$Homer L. Dodge Department of Physics and Astronomy, University of Oklahoma, Norman OK; United States of America.\\
% ...
$^{137}$$^{(a)}$Departamento de F\'isica, Pontificia Universidad Cat\'olica de Chile, Santiago;$^{(b)}$Millennium Institute for Subatomic physics at high energy frontier (SAPHIR), Santiago;$^{(c)}$Instituto de Investigaci\'on Multidisciplinario en Ciencia y Tecnolog\'ia, y Departamento de F\'isica, Universidad de La Serena;$^{(d)}$Universidad Andres Bello, Department of Physics, Santiago;$^{(e)}$Instituto de Alta Investigaci\'on, Universidad de Tarapac\'a, Arica;$^{(f)}$Departamento de F\'isica, Universidad T\'ecnica Federico Santa Mar\'ia, Valpara\'iso; Chile.\\
% ...
$^{150}$Department of Physics, Technion, Israel Institute of Technology, Haifa; Israel.\\
$^{151}$Raymond and Beverly Sackler School of Physics and Astronomy, Tel Aviv University, Tel Aviv; Israel.\\
% ...
$^{a}$ Also Affiliated with an institute covered by a cooperation agreement with CERN.\\
$^{b}$ Also at An-Najah National University, Nablus; Palestine.\\
$^{c}$ Also at APC, Universit\'e Paris Cit\'e, CNRS/IN2P3, Paris; France.\\
$^{d}$ Also at Borough of Manhattan Community College, City University of New York, New York NY; United States of America.\\
$^{e}$ Also at Center for High Energy Physics, Peking University; China.\\
$^{f}$ Also at Center for Interdisciplinary Research and Innovation (CIRI-AUTH), Thessaloniki; Greece.\\
% ...

\end{flushleft}
